\documentclass[useAMS,usenatbib]{mn2e}
\usepackage{ upgreek } %me
\usepackage{graphicx}
\usepackage{caption}
\usepackage{subcaption}
\usepackage{amssymb}
\usepackage{amsmath}
\usepackage{float}
\usepackage{color, array}
\usepackage{booktabs} % Required for better horizontal rules in tables
\usepackage{lscape}

\title[Measuring Limb Darkening of Stars in high magnification Microlensing event with FEM ]{Measuring Limb Darkening of Stars in high magnification Microlensing Events by the Finite Element Method}
\author[Golchin ‌and Rahvar]{L. ~Golchin$^{1}$\thanks{E-mail: lgolchin@physics.sharif.edu}, S. ~Rahvar$^{1}$\thanks{E-mail: rahvar@sharif.edu} \\
$^{1}$Physics Department, Sharif University, P.O.Box 11365-9161, Azadi Avenue, Tehran, Iran}

\pagerange{\pageref{firstpage}--\pageref{lastpage}} \pubyear{2016}

\def\LaTeX{L\kern-.36em\raise.3ex\hbox{a}\kern-.15em
T\kern-.1667em\lower.7ex\hbox{E}\kern-.125emX}

\begin{document}
\label{firstpage}

\maketitle

\begin{abstract}
The finite-size effect in gravitational microlensing provides a possibility to measure the limb darkening of distant stars. We use the Finite Element Method (FEM) as an inversion tool for discretization and inversion of the magnification-limb darkening integral equation. This method makes no explicit assumption about the shape of the brightness profile more than the flatness of the profile near the center of the stellar disk. From the simulation, we investigate the accuracy and stability of this method and we use regularization techniques to stabilize it. Finally, we apply this method to the single lens, high magnification transit events of OGLE-2004-BLG-254 (SAAO-I), MOA-2007-BLG-233/OGLE-2007-BLG-302 (OGLE-I, MOA-R), MOA-2010-BLG-436 (MOA-R), MOA-2011-BLG-93 (Canopus-V), MOA-2011-BLG-300/OGLE-2011-BLG-0990 (Pico-I) and MOA-2011-BLG-325/OGLE-2011-BLG-1101 (LT-I) in which light curves have been observed with a high cadence near the peak \citep{Ch}. The recovered intensity profile of stars from our analysis for five light curves are consistent with the linear limb darkening and two events with the square-root profiles. The advantage of FEM is to extract limb darkening of stars without any assumption about the limb darkening model.
\end{abstract}


\section{Introduction}
The limb of stellar disks is dimmer and redder than their center, this effect is known as Limb Darkening (LD) effect. LD happens because photons coming from the center of the stellar disk originate deeper in the photosphere than photons from the limb where the temperature is lower \citep{Gr}. The result is that the light from the center of the stellar disk is more intense and its temperature is higher. Direct study of stellar LD is possible by interferometric photometry of nearby giant stars in one or several bands \citep{Bu,Per,Auf,Wit,Mnt}. The other method is studying special events such as eclipsing binaries \citep{Pop,Sou,Sou2}  or occulting systems \citep{Ric}. 

 The other new method is the finite-size effect in the gravitational microlensing events \citep{Alb,An,Yoo,Ch,R15} which is the subject of our study and can be used as a probe to scan the intensity profile of distant stars. Gravitational microlensing happens when a massive astronomical object inside the Milky Galaxy intervenes as a background star and bends its light toward the observer. Since the observer, lens and source are moving inside the Galaxy, the angular position of the lens compared to the source changes by time and as the angular separation gets closer, it results in the increase of the magnification of the source star \citep{Paz1986}. The time-scale of magnification scales with the square root of the lens mass and can take from hours to almost one month.

According to the original paper by Einstein (1936), he investigated the gravitational lensing of a background star by another star and he stated that "there is no great chance of observing this phenomenon". This phenomenon has also investigated by pioneer astronomers such as F. Link. For more details refer to \citet{DV}. Due to instrumental progress, nowadays thousands of microlensing events are observed towards the center of Galaxy by OGLE\footnote{http://ogle.astrouw.edu.pl/} and MOA\footnote{http://www.phys.canterbury.ac.nz/moa/} surveys and other follow-up groups as $\mu$-Fun\footnote{http://www.astronomy.ohio-state.edu/~microfun/}, MindStep\footnote{http://www.mindstep-science.org/}, Planet\footnote{http://www.planet-legacy.org/}. The gravitational microlensing has broad astrophysics applications such as investigating dark compact objects so-called MACHOs \footnote{Massive Astrophysical Compact Halo Objects} in the Milky Way halo \citep{Paz1986}. The two observational groups of EROS and MACHO after a decade monitoring of Magellanic Clouds for the microlensing events concluded that MACHOs don't have a significant contribution in the dark matter contribution of the Galactic halo  \citep{Al2000, Af2003}. 

The other application of microlensing is using this method for detecting extrasolar planets \citep{Ga08,Ga12,Ts18} and even detecting signals from Extraterrestrial intelligent life \citep{Ra16}. Also, the microlensing can be used for studying the stellar spots on the source star by polarimetry \citep{SWN95, S15} and time variation of center of light of the source star by astrometry \citep{Wa95, SR15}. Studying the structure of  Milky Way through the combination of photometry and astrometry observations with GAIA is another important application of gravitational microlensing \citep{RG05, Mnz}.

In this work, our aim is the application of gravitational microlensing for studying the LD of the source stars during the lensing. In microlensing events with the minimum impact parameter comparable with the size of the source star, the source star cannot be taken as a point-like object. The result of this effect, so-called finite-size effect \citep{sch86, sch87, Wi94} is the deviation of the light curve from a point-like source around the peak. The other feature of this effect is that when the lens crosses over the source star, the main contribution of light is received from the location of the lens on the source star. This effect turns the microlensing effect to an astronomical scanner that can probe the surface of the source stars. This kind of source scanning also happens in the binary lenses where the lensing system produces caustic lines. The observation of these events with high cadence allows us to probe the detailed structure of the source star such as LD  \citep{DV95,Wi95,DV98,Fie03, Ca06} and stellar spots \citep{He2000, Hen02}.  

Here, we study the single-lens finite-size effect in the microlensing events to recover the LD of the source star, using the Finite Element Method (FEM) \citep{Zie}. This method is an inversion tool to numerically solve the magnification-LD equation. There are other inverse methods that have been used for recovering limb darkening of source stars such as discretization \citep{Ga, Bog}, Backus-Gilbert method \citep{GC01} and PCA \footnote{Principal Component Analysis} inversion \citep{He}. \citet{He} presents a detailed review about numerical methods for recovering LD.  The magnification-LD equation is a Fredholm integral equation of the first-kind \citep{Waz11} and the result of solving this equation is recovering the limb-darkening profile data from the light curve data around the peak. 

In section (\ref{gm}), we briefly introduce gravitational microlensing and finite-size effect. In section (\ref{section2}) we explain the FEM approach and apply it to a generic Fredholm integral equation of the first-kind. In section (\ref{simData}) we apply FEM to the magnification-LD equation by suitable adjustment of stellar disk mesh and adequate numerical integration technique. To examine systematic errors we recover the brightness profile from non-noisy simulated light curves; then to examine the accuracy and stability of our inversion method, we add different ranges of Gaussian noise to flux values and estimate the variation of LD profile induced by the flux noise. To achieve stability and better accuracy we introduce a data selection algorithm and use regularization techniques on FEM solutions. Finally, in this section, we apply our method to the single lens, high magnification data ofOGLE-2004-BLG-254 (SAAO-I\footnote{South African Astronomical Observatory, South Africa, I ~passband}), MOA-2007-BLG-233/OGLE-2007-BLG-302 (OGLE-I\footnote{Las CampanasObservatory, Chile}, MOA-R\footnote{Mt. John Observatory,NewZealand}), MOA-2010-BLG-436 (MOA-R), MOA-2011-BLG-93 (Canopus-V\footnote{Canopus Hill Observatory}), MOA-2011-BLG-300/OGLE-2011-BLG-0990 (Pico-I\footnote{Observatorio do Pico dos Dias, Brazil}) and MOA-2011-BLG-325/OGLE-2011-BLG-1101 (LT-I\footnote{Liverpool Telescope, La Palma, Spain}) events to extract directly the LD profile of the source star. The conclusion is given in section (\ref{conc}). 

\section{Gravitational Microlensing and Finite Size Effect}
\label{gm}
 
 When the light ray of a star (source) passes closely enough to another astronomical object (lens) it bends due to the gravitational field of the lens \citep{Ei36}. This effect causes secondary images from the source \citep{Ed20} or a ring image in the case that we have perfect alignment of the source, lens, and observer \citep{Ch24}. The angular radius of this ring is called angular Einstein radius,   $$\theta_E = \sqrt{\frac{4GM}{c^2}\frac{D_{ls}}{D_lD_S}}.$$  If the separation of resultant images is of the order of micro arcsec (microlensing events) the images are not resolvable but the source star will be magnified \citep{Paz}. During a microlensing event, the apparent brightness of the source star will rise and finally drops to the baseline.  The time-dependent magnification of a point-like source by a single lens is as follows:
\begin{equation}\label{eq:Apoint}
 A(u) = \frac{u^2+2}{u\sqrt{u^2+4}},\,\, u=(u_0^2+\frac{(t-t_0)^2}{t_E^2})^{\frac{1}{2}}
 \end{equation}
 in which $u$ is the angular separation of the lens and source in units of $\theta_E$, 
 $u_0$ is the minimum impact parameter, $t_0$ is the time of maximum magnification and $t_E$ is the Einstein timescale, corresponds to the time-scale that source move an angular distance of $\theta_E$ .

 For the case that the minimum impact parameter (i.e. $u_0$) becomes comparable with the angular size of source star (normalized to the Einstein angle),
 the lens amplifies different parts of the source star with different weights and it makes possible to study the surface brightness profile and size of the source star \citep{Wi95}. In this case, the magnification is obtained from the convolution of equation (\ref{eq:Apoint}) and surface brightness of the source. In the observations, the high magnification events alerted well before the peak so that a network of follow-up telescopes can perform high cadence observation \citep{Al97, Ch}. From the measurement of the light curve around the peak, one can calculate the angular size of the source, $\theta_\star$ in units of $\theta_E$ (i.e. $\rho_{\ast} = \theta_\star/\theta_E$). Knowing the type of the source star and the distance of the source star from the observer which is mainly located at the Galactic Bulge, one can measure the Einstein angle of the lens.  
  
The second channel for the finite-size effect observation is during the caustic crossing of the binary lens where the LD also can be measured with this method \citep{Al99, Zub}.In this paper, our aim is to study the single-lens very high magnification events where due to a large number of microlensing events, the number of these types of events has been increased in recent years. Such events have been reported by \citet{Ch} where standard linear Limb Darkening Coefficient (LDC) of source stars have been obtained from the fitting model with the light curve. 

Throughout this paper we adapt the normalized impact parameter to the angular size of star by $\ell={u}/{ \rho_{\ast} }$, normalized minimum impact parameter to the size of star by $\;p={u_0}/{\rho_{\ast}}$ and the transit time scale of lens crossing over the star by $\;t_{\ast}= \rho_{\ast}t_E$. Here, $\ell=1$ corresponds to when the lens enters or leaves the source disk. Now we can calculate the magnification of a source with circular symmetric LD profile (i.e. $I(r)$) from a single lens  as follows:
\begin{eqnarray}\label{eq:Ft}
 A(\ell) &=& \frac{F(\ell)}{F_0}=\frac{1}{F_0} \int_0^1 \, A(\ell,r) \, I(r)\, r \mathrm{d}r,
 \end{eqnarray}
where $r$ is within the range of $[0,1]$ and $\ell$ depends on the location of lens with respect to the centre of source star and is given by  
\begin{equation}
\label{ell}
\ell=\sqrt{p^2+\frac{(t-t_0)^2}{t_{\ast}^2}}, 
\end{equation}
$F_0=2\pi\int_0^1 \, I(r)\,r \mathrm{d}r$ and $A(\ell,r)$ is the  angle-integrated amplification:
\begin{eqnarray} \label{eq:A}
 A(\ell,r)&=&\int_0^{2\pi}A(\ell,r,\phi)\mathrm{d}\phi,  \\ \nonumber \text{and} \\
 A(\ell,r,\phi)&=&\frac{x^2+2\epsilon^2}{x\sqrt{x^2+4\epsilon^2}}\;\;,
 x^2=\ell^2+r^2-2 \ell r\cos{\phi}. 
\end{eqnarray}
where  $\epsilon={1}/{\rho_{\ast}}$. Equation (\ref{eq:A}) can be written  in terms of elliptic integrals as follows \citep{G00} :
\begin{eqnarray} \label{eq:Alr} 
 A(\ell,r) &=& \frac{4}{(\ell+r) \sqrt{(\ell-r)^2+4\epsilon^2}}\times  [2\epsilon^2 K(\kappa) \\ \nonumber
 &+&(\ell-r)^2\Pi(\alpha^2,\kappa)],\\ \nonumber
 {\text where } \\		
 \kappa&=& \frac{4\epsilon}{\ell+r} \sqrt{\frac{lr}{(\ell-r)^2+4\epsilon^2}} ,\; \alpha^2=-\frac{4lr}{(\ell+r)^2}. \nonumber
\end{eqnarray}
 $K$ and $\Pi$ are the first and the third-kind elliptic integrals, as follows:
 
 \begin{eqnarray}
 K(\kappa)&=&\int_0^{\frac{\pi}{2}} \frac{ \mathrm{d}\theta}{ \sqrt{1-\kappa^2 sin^2\theta}}, \nonumber \\
 \Pi(\alpha^2,\kappa)&=&\int_0^{\frac{\pi}{2}} \frac{ \mathrm{d}\theta}{(1+\alpha^2 sin^2 \theta ) \sqrt{1-\kappa^2 sin^2\theta}}.
 \end{eqnarray}

\begin{figure}
\centering
  \includegraphics[width=90mm]{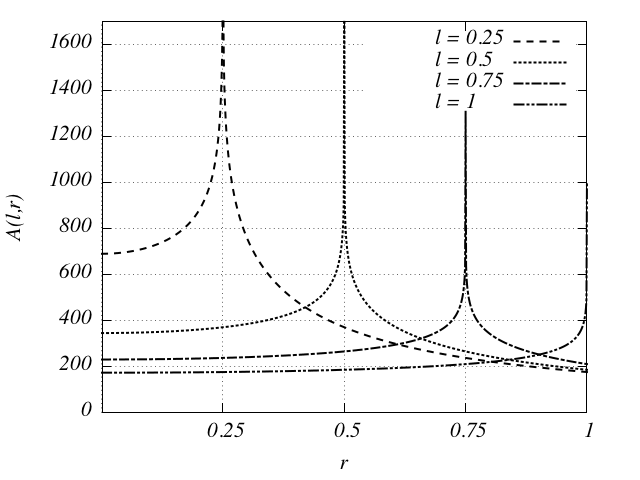}
  \caption{$A(\ell,r)$ has logarithmic divergency at $r=\ell$. In this figure $\epsilon = 27.47$ and $p=0.1648$.}
  \label{fig:IA}
\end{figure}

  $A(\ell,r)$ has a logarithmic divergence in $r=\ell$ as shown in Figure (\ref{fig:IA}), meaning thereby that at $r \to \ell$, parts of source star close to this point are amplified much more than the other parts. This property turns microlensing to a natural surface scanner. 
  %Also using this property as we will discuss in Section (4), allows us to estimate equation (\ref{eq:Ft}) by a well-posed set of algebraic equations in FEM.
The angle-integrated amplification can also be approximated near this divergency as follows \citep{He}:
\begin{eqnarray}  \label{eq:aprox} 
 A(\ell,\ell+\delta) &=& \frac{2\epsilon}{\ell}\left(1-\frac{\delta}{2l}\right) \ln{\frac{8\epsilon \ell}{|\delta|\sqrt{\ell^2+\epsilon^2}}}+ \\ \nonumber 
 &\;&4\arctan{\frac{\ell}{\epsilon}}+ \frac{\epsilon(2l^2+\epsilon^2)}{\ell^2(\ell^2+\epsilon^2)}\delta+\mathcal{O}(\delta^2\ln{|\delta|}).
\end{eqnarray}

\section{Finite Element Method in Fredholm Integral Equations of the first-kind}
\label{section2}

The equation (\ref{eq:Ft}) as the magnification-LD integral is a Fredholm integral equation of the first-kind. We deal with such equations in several other situations in the astronomy \citep{CB86}. As the history of application of this method in astronomy, that was used in the galactic dynamics for modeling perturbed stellar systems \citep{Ja} and  constructing smooth distribution functions of stellar systems \citep{JT}. Let us take the integral as follows  \begin{equation} \label{eq:Fred}g(t)=\int_{x_b}^{x_f}K(x,t)f(x)dx,\end{equation}in which $f(x)$ is unknown and $K(x,t)$ and $g(t)$ are known. Let us first introduce the Product Integration Method (PIM) which is simpler than FEM but similar in some aspects, in PIM one takes $N$ data points : $(t_i, g(t_i))$ and divides the $x$-domain into $M$ parts ($M\geqslant N$)and  writes discrete version of equation(\ref{eq:Fred}) as follows:\begin{equation} \label{eq:PIM}g(t_i)=\sum^M_{j=1}{\int_{x_j}^{x_{j+1}}K(x,t_i)f(x)dx},\;\; i=1..N\end{equation}Then by choosing a $f(x)$ to be piecewise constant or piecewise linear over each part, a simple algebraic formula can be derived for each data point.Then one can gather all above $N$ equations into a $N\times M$ set of algebraic equations.If one takes $M>N$ additional constraints such as monotonically and positiveness can be met. See \citet{CB86} for more details.The main difference between FEM and PIM is that in FEM we assure the continuity of the solution and we can estimate data as a continuous piecewise polynomial as well. Below we explain this in more detail.

\begin{figure}
\centering
  \includegraphics[width=65mm]{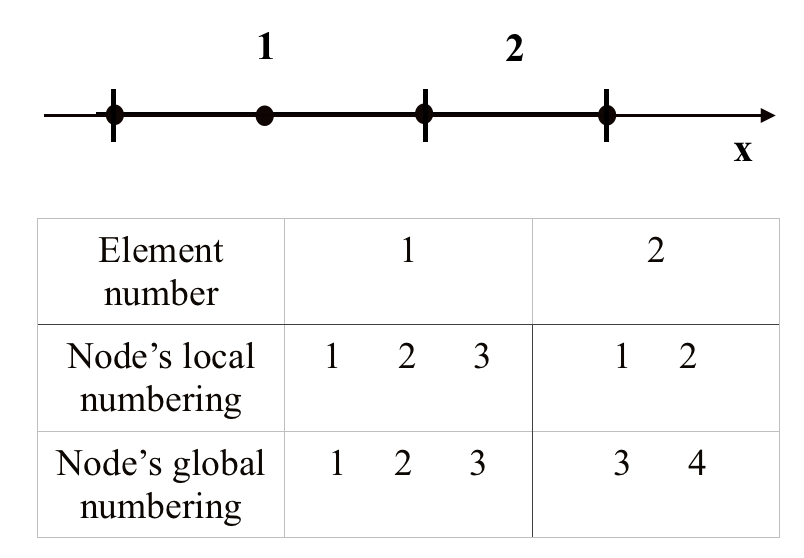}
  \caption{ A simple example of FEM-mesh with 2 elements, 4 nodes and their numbering. Notice that element sizes and number of their nodes could be different.}
  \label{fig:nmbrng}
\end{figure}

To solve equation (\ref{eq:Ft}) by use of FEM, we approximate $g(t)$ as a continuous piecewise $n$-degree polynomial and we find $f(x)$ as a continuous piecewise $m$-degree polynomial. To do this, first we divide $x$-domain into $M$ elements, each element contains $n_{d2}=m+1$ nodes on its boundaries or in its interior. Then we give two numbers to each node, one indicates the location of the node in the element (local numbering) and the other one is the location in the whole domain (global numbering), the total number of nodes $N_2$ is $N_2=M(n_{d2}-1)+1$.  See  Figure (\ref{fig:nmbrng}) as an example of a one-dimensional FEM-mesh and its numbering method where the size of elements essentially are not equal. 

  We can use certain local coordinates within each element, so that ranges of all elements become the same: $x_i^1\le x \le x_i^{n_{d2}} \to \, -1\le \bar{x} \le 1$. These local coordinates are defined as bellow:
\begin{equation}\label{eq:bar}
	\bar{x}=2\frac{x-x_j^1}{\Delta_j}-1,\;\;
	\Delta_j=x_j^{n_{d2}}-x_j^1.
\end{equation}

\begin{figure}
\centering
  \includegraphics[width=60mm]{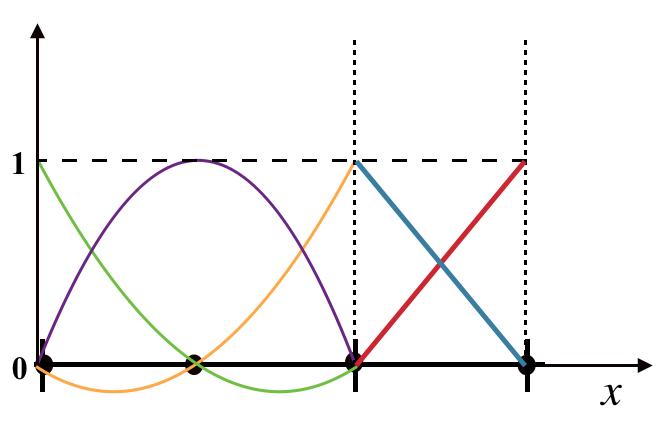}
  \caption{FEM-mesh with 2 elements, 4 nodes. First element is 2nd-order and has three shape functions. Second element is 1st-order and has two shape functions.}
  \label{fig:shps} 
\end{figure}

Within each element we write $f(x)$ as a linear combination of $n_{d2}$  functions which are $m$-degree polynomials. We choose these polynomials such that coefficients of the linear combination become nodal values ($f_j^{j_d}$, in which $j$ is the element number and $j_d$ is the local node number.). Therefore each polynomial should take the value of 1 in one node and value of 0 in other nodes, these basis functions are called shape functions in the FEM context (See Figure (\ref{fig:shps})). 

Hence we can write the approximation of $f(x)$ within $j$th  element as follows:
\begin{equation}\label{eq:combf}
f_j(\bar{x})=\sum_{j_d=1}^{n_{d2}}U_j^{j_d}(\bar{x})f_j^{j_d},\;\; j=1..M,
\end{equation}

\begin{figure}
\centering
  \includegraphics[width=65mm]{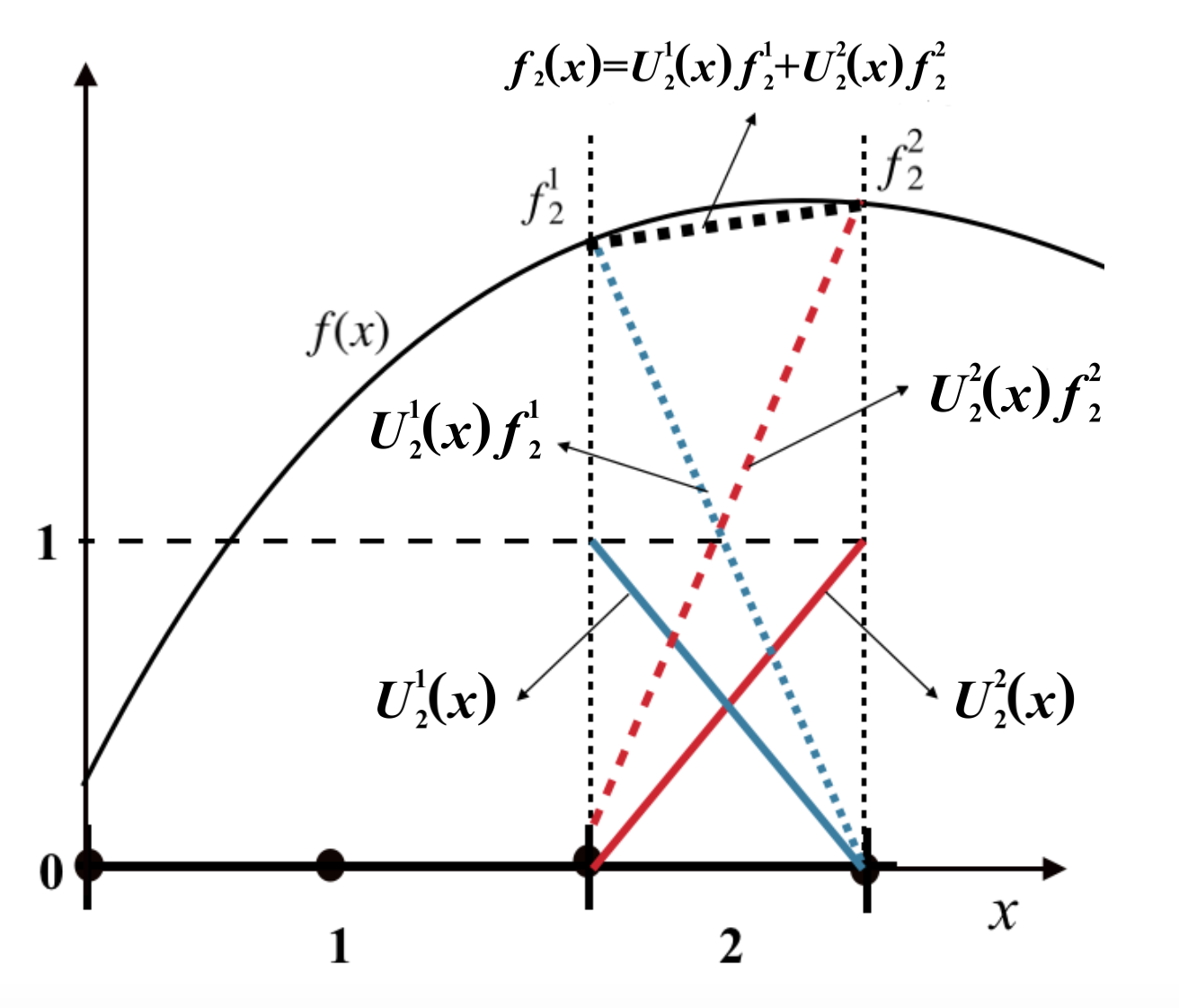}
  \caption{FEM-mesh with 2 elements, approximation of $f(x)$ within second element using shape functions ($U_2^1(x), U_2^2(x)$).}
  \label{fig:aprx}
\end{figure}
where $U_j^{j_d}$s as seen in Figure (\ref{fig:aprx}) are the shape functions. We can write equation (\ref{eq:combf}) in a compact form as a vector inner product (throughout this article we denote vectors and matrices by $\mathbf{bold}$ characters, dot product by  $.$, transpose by $^T$ and $\otimes$ stands for the dyadic product.):
  \begin{eqnarray}\label{eq:dscrt1}
	f_j(\bar{x})&=&\mathbf{U}_j(\bar{x})\,.\,\mathbf{f}_j \\
	\mathbf{U}_j&=&( U_j^1(\bar{x})\;\;...\;\;U_j^{n_{d2}}(\bar{x})),  \;\; \mathbf{f}_j=( f_j^1\;\; ...\;\;f_j^{n_{d2}})^T
\end{eqnarray}

Now we can write piecewise approximation of $f(x)$ in whole $x$-domain by  summing over $f_j(\bar{x})$ of all elements:
\begin{equation}
	\tilde{f}(x)=\sum_{j=1}^M H_j(x)\,f_j(\bar{x}), \,\label{sumf}
\end{equation}
where $H_j(x)$ is the top hat function where it is zero every where and one within $j$th element.

Let us go back to equation (\ref{eq:Fred}), if $g(t)$ is known in $N_1$ points in $t$-domain, we can write its approximation as a piecewise $n$-degree polynomial by the same procedure:
  \begin{eqnarray}\label{eq:dscrt2}\label{dotg}
  \tilde{g}(t)&=&\sum_{i=1}^N H_i(t)\,g_i(\bar{t}), \;\;
	g_i(\bar{t})=\mathbf{V}_i(\bar{t})\,.\,\mathbf{g}_i ,\\  
	N&=&\frac{N_1-1}{n_{d1}-1}, n_{d1}=n+1\\ \nonumber
	\mathbf{V}_i(\bar{t})&=&( V_i^1(\bar{t})\;\;...\;\;V_i^{n_{d1}}(\bar{t})),   \;\; \mathbf{g}_i=(g_i^1\;\;...\;\;g_i^{n_{d1}})^T
\end{eqnarray}

By substituting (\ref{dotg}) and (\ref{sumf}) in (\ref{eq:Fred}), in the FEM formalism this equation can be written as: 
\begin{equation}\label{eq:gt}
 \sum_{k=1}^N H_k(t)\,\,g_k(\bar{t}) = \sum_{j=1}^M\int_{x_b}^{x_f}H_j(x)K(x,t)f_j(\bar{x})dx .
 \end{equation}
 If we multiply both sides of the above equation by $H_i(t)$ we get an equation for the $i$th element:
\begin{eqnarray}\label{eq:gtnm}
&&H_i(t)\,\mathbf{V}_i(\bar{t})\,.\,\mathbf{g}_i = \nonumber \\
&&H_i(t)\sum_{j=1}^M\int_{x_b}^{x_f}H_j(x)K(x,t)\mathbf{U}_j(\bar{x})dx.\mathbf{f}_j,\;i=1..N .
 \end{eqnarray} 

 The above equation shows the relation between $i$th element of $t$-domain and all elements of $x$-domain, hence we can proceed to derive a relation between nodal values of both sides. To do so we use Galerkin projection technique \citep{Zie}: we operate (\ref{eq:gtnm}) with $dt \mathbf{U}_i \otimes$.Then we integrate the result over the $t$-domain and transform to the local coordinate systems of $t$ and $x$-domains, we get:

\begin{eqnarray}\label{eq:femg}
&&\mathbf{g}_i =\mathbf{G}_{nn}^{-1}.\sum_{j=1}^M \frac{\Delta_j}{2} \int_{-1}^{1}\int_{-1}^{1}K(x,t)\mathbf{V}_i(\bar{t}) \otimes\mathbf{U}_j(\bar{x})d\bar{x} \bar{t}\,.\,\mathbf{f}_j ,\nonumber  \\
&& \text{where}  \\
&&\mathbf{G}_{nn}=\int_{-1}^{1} \mathbf{V}_n(\bar{t}) \otimes\mathbf{V}_n(\bar{t})d\bar{t}\,.
\end{eqnarray}

We can write the right hand side of (\ref{eq:femg}) in terms of a summation over $M$,  $n_{d1}\times n_{d2}$ matrices ($\mathbf{E}_{ij}$): 

\begin{eqnarray} \label{eq:Eij} 
\mathbf{g}_i&=&\sum_{j=1}^{M} \mathbf{E}_{ij}.\mathbf{f}_j ,\;\; i=1..N,\\ \text{where} \nonumber \\
 \mathbf{E}_{ij}&=&\frac{\Delta_j}{2}\mathbf{G}_{nn}^{-1}.\int_{-1}^1\int_{-1}^1 K(x,t)\mathbf{V}_i(\bar{t})\otimes \mathbf{U}_j(\bar{x})\,d\bar{x} d\bar{t} \nonumber
 \end{eqnarray}
 
\begin{figure}
\centering
  \includegraphics[width=55mm]{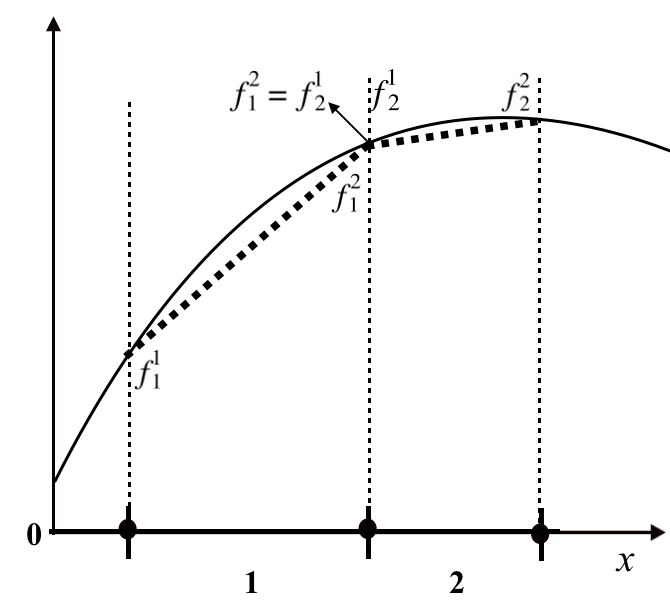}
  \caption{Continuity condition of the FEM piecewise approximation of $f(x)$.}
  \label{fig:cnty}
\end{figure}

Next step is assembling all $N$ set of equations (\ref{eq:Eij}) into one large set of algebraic equations, considering equality of nodal values in common nodes of neighbour elements ($g_i^1=g_{i-1}^{n_{d1}},\;f_j^1=f_{j-1}^{n_{d2}}$), see Figure (\ref{fig:cnty})).

\begin{figure}  
\centering

   \begin{subfigure}{.4\textwidth}
   \caption{}
   \includegraphics[width=70mm]{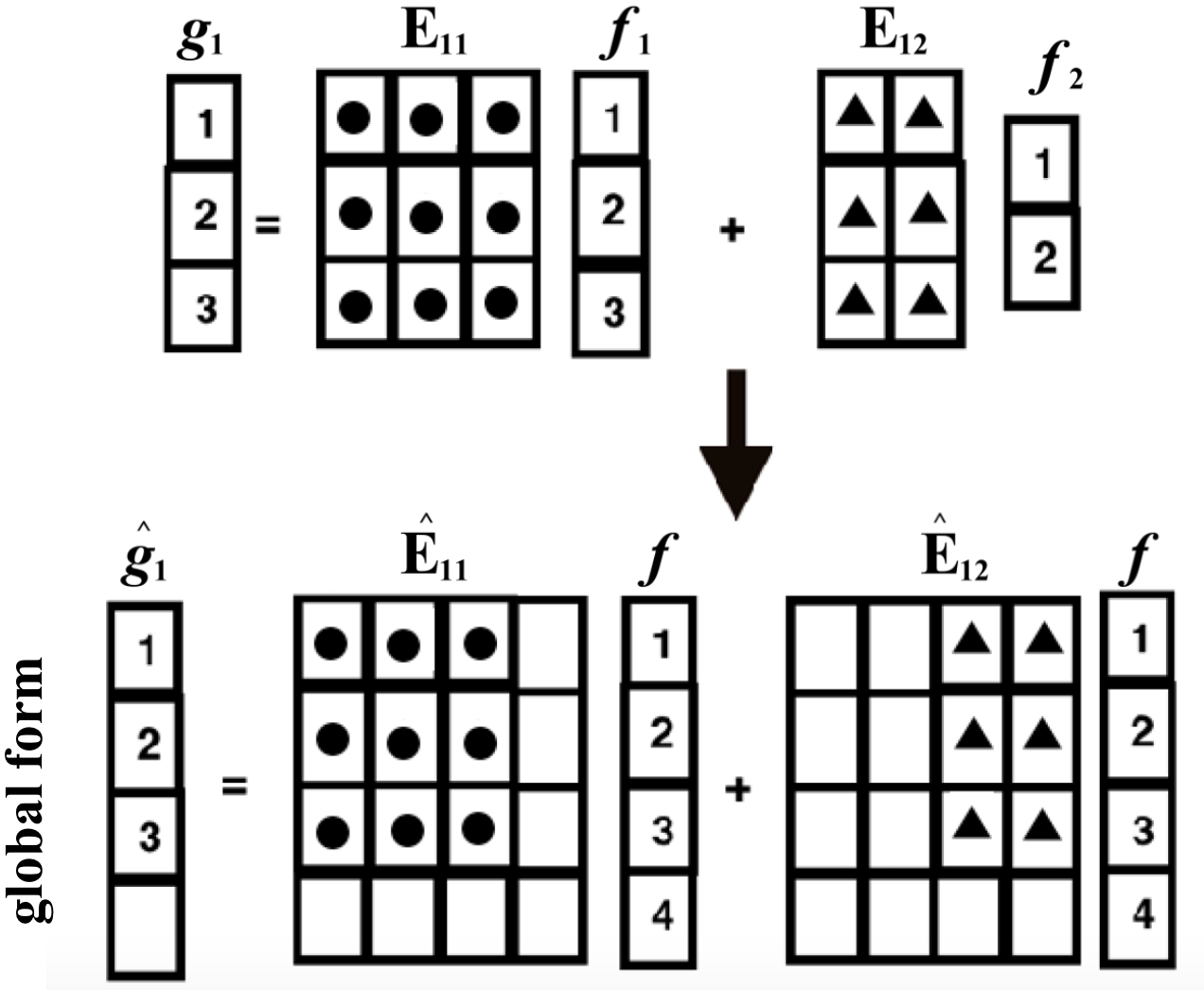}
   \label{subfig:g1}
   \end{subfigure}
  
   \begin{subfigure}{.4\textwidth}
   \caption{}
   \includegraphics[width=70mm]{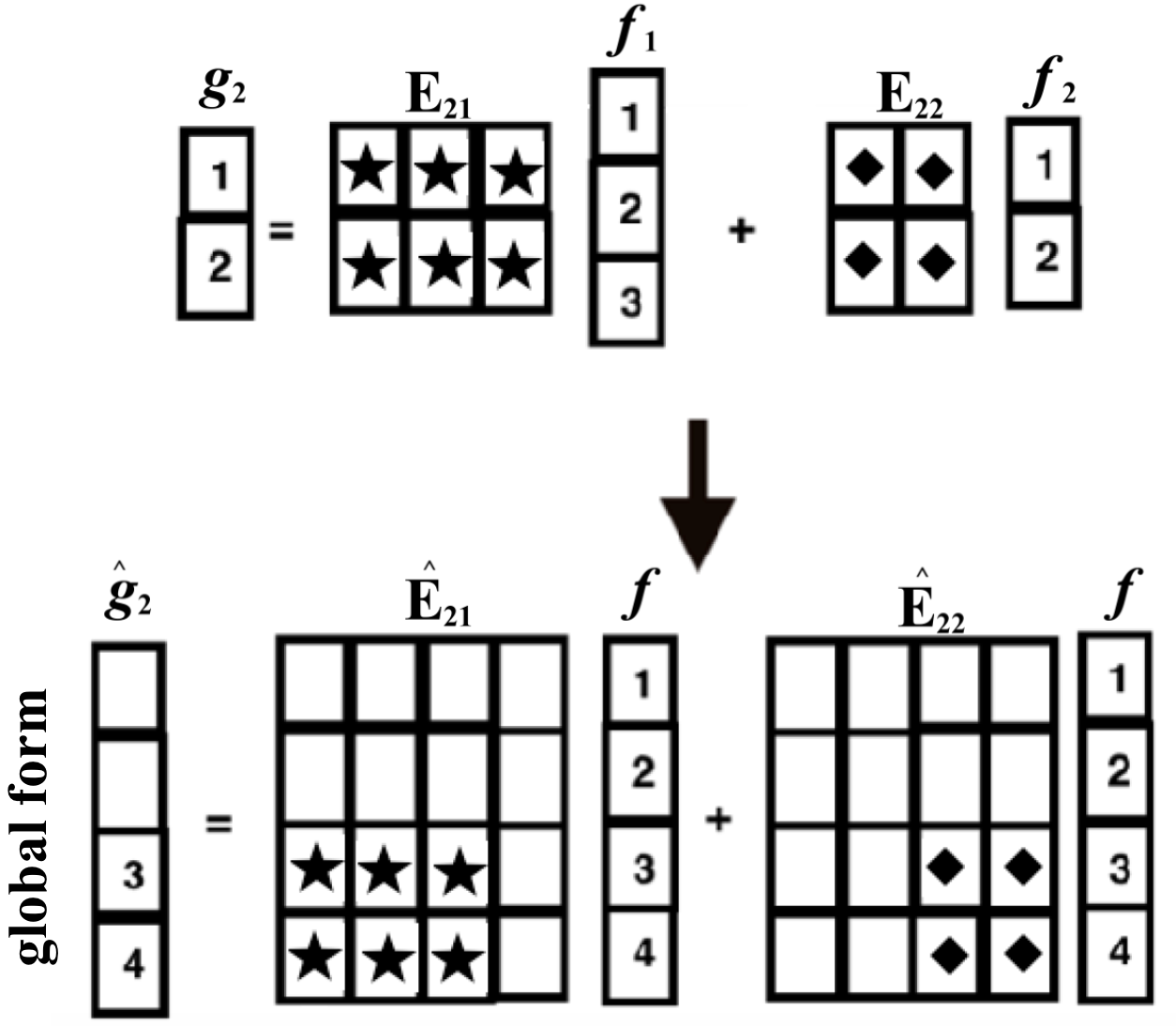}
   \label{subfig:g2}
   \end{subfigure}
  
\caption{Schematic presentation of transformation of local element matrices into global forms for equation \ref{Aisum}, assuming the mesh of figure \ref{fig:nmbrng} for $x$ and $t$ domains. (a) first element of $t$-domain, (b)second element of $t$-domain. }
\label{tglb}
\end{figure}

\begin{figure}
\centering
  \includegraphics[width=70mm]{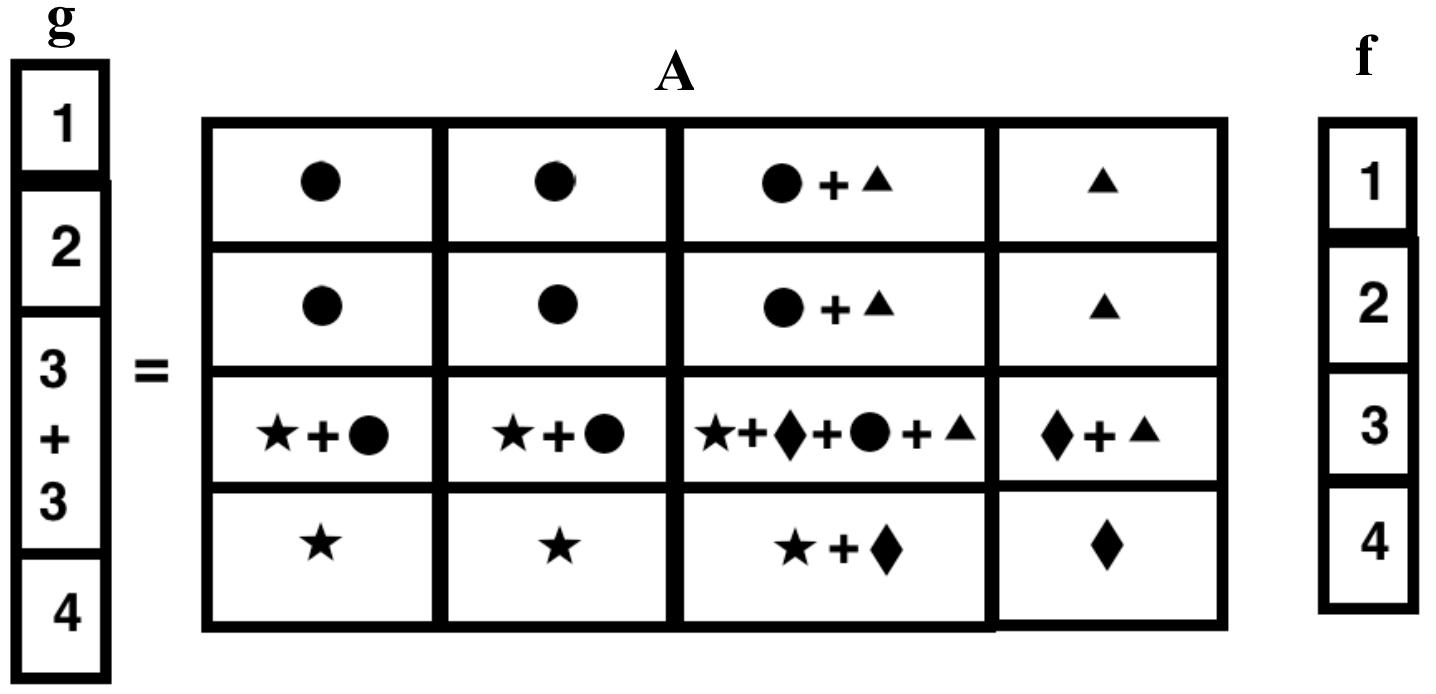}
  \caption{Schematic presentation of assembling process (equation \ref{eq:GSM}) for global forms of equations \ref{Eij} shown in figure \ref{tglb}.} 
  \label{fig:asmbl}
\end{figure}

 To do this we use global numbering of nodes, we rewrite all $\mathbf{E}_{ij}$ matrices which have the $n_{d1}\times n_{d2}$ dimension in form of $N_1\times N_2$ matrices (i.e. $\mathbf{\hat{E}}_{ij}$) and $\mathbf{g}_i$ in form of a $N1 \times 1 $ vector (i.e. $\mathbf{\hat{g}}_i$). In other words, $ E_{ij}^{i_d j_d} \to \hat{E}_{ij}^{IJ}$ and  $g_i^{i_d} \to \hat{g}_i^I $ where $( i_d=1..n_{d1}, j_d =1..n_{d2}, I=1..N_1, J=1..N_2)$.\\
  This process is shown in a schematic way in Figure (\ref{tglb}).  We get $N$ set of algebraic equations,
   we assemble them all to get a global set of equations which is the approximation of the original integral-equation (\ref{eq:Fred}):
 \begin{eqnarray}\label{eq:GSM}
\mathbf{g}=\mathbf{A}.\mathbf{f},\;\; \mathbf{A}=\sum_{j=1}^M\sum_{i=1}^N \mathbf{ \hat{E}_{ij} },\;\; \mathbf{g}=\sum_{i=1}^N\mathbf{\hat{g}_i},
\end{eqnarray}  
 
  in which $\mathbf{f}$ is the vector of nodal values of $f(x)$ ordered by their global numbering and the $(N_1\times N_2)$ matrix ($ \mathbf{A}$) is called Global  Stiffness Matrix (GSM) (see Figure (\ref{fig:asmbl}) for a schematic description of assembling process).
   
Now if we get a well-posed GSM we can derive nodal values of $f(x)$ by solving linear algebraic equation (\ref{eq:GSM}) otherwise one could use regularization techniques described by \citet{CB86}.  Then using (\ref{sumf}), the piecewise approximation of $f(x)$ can be calculated in entire $x$-domain.

  \section{FINITE ELEMENT METHOD IN MICROLENSING}
\label{simData}
In this section, we apply FEM in the high magnification microlensing event, with the finite-size effect to recover the LD profile (the integral-equation \ref{eq:Ft}), following the procedure described in the former section. Suppose we have $A(\ell)$ in $N_1$ nodes, so we can divide the $\ell$-space into $N$ elements and write the Continuous Piecewise Polynomial (CPP) approximation of $A(\ell)$ by using shape functions that we introduced before as follows:
\begin{eqnarray}
\label{Apw}
A(\ell)=\sum_i^{N}H_i\,U_i(\bar{\ell}).A_i, \; \bar{\ell}=2\frac{\ell-\ell_{i}}{\ell_{i+1}-\ell_i}-1
\end{eqnarray}
 
 We impose the error in each magnification as $A_i=A(l_i)\pm \delta A_i$
where $l_i={u(t_i)}/{\rho_{\ast}},\,i=1..N_1$.We write $1$-degree CPP approximation of normalised LD profile (i.e. $I(r)/{F_0}$) by total $N_2$ nodes and $M=N_2-1$ elements in $r$-space ($r_1, r_2, ..., r_{N_2}$) :

\begin{eqnarray}
\label{Ipw}
I(r)=\sum_j^{M}H_j\,V_j(\bar{r}).I_j, \; \bar{r}=2\frac{r-r_i}{r_{i+1}-r_i}-1
\end{eqnarray}

From equation (\ref{eq:Eij}) using notation for gravitational microlensing, we rewrite these equations as follows:

\begin{eqnarray} 
\label{Eijs} 
\mathbf{A}_i&=&\sum_{j=1}^{M} \mathbf{E}_{ij}.\mathbf{I}_j, \,\,  i=1.. N ,  \label{Aisum}  \\  \nonumber
{\text where:}\\ \nonumber
 \mathbf{E}_{ij}&=&\frac{\Delta_j}{2}\mathbf{G}_{ii}^{-1}.\int_{-1}^1\int_{-1}^1 r\,A(r,\ell)\mathbf{U}_i(\bar{\ell})\otimes \mathbf{V}_j(\bar{r})\,d\bar{\ell} d\bar{r} , \label{Eij} \\  \nonumber
\mathbf{G}_{ii}&=&\int_{-1}^{1} \mathbf{U}_i(\bar{\ell}) \otimes\mathbf{U}_i(\bar{\ell})d\bar{\ell}. \nonumber
 \end{eqnarray}
 
$\mathbf{E}_{ij}$s in equation (\ref{Eijs}) due to logarithmic divergency of $A(l_i,r)$ at $r=l_i$ (equation \ref{eq:aprox}), can not be calculated by simple numeric integration methods . To carry out $\bar{r}$ integrations we use Runge-Kutta adaptive step size method \citep{Pr}. With this method we calculate the $\bar{r}$ integrations up to precision of $\mathcal{O}(10^{-7})$. We use Gaussian quadrature for $\bar{\ell}$ integration. By assembling $\mathbf{E}_{ij}$s matrices we get the global algebra set of equations which is the approximation of the original integral-equation (\ref{eq:Ft}):

 \begin{eqnarray}\label{eq:FtFEM}
\mathbf{A}=\mathbf{M}.\mathbf{I},\;\; \mathbf{M}=\sum_{j=1}^M\sum_{i=1}^N \mathbf{ \hat{E}_{ij} },\;\; \mathbf{A}=\sum_{j=1}^M\mathbf{A}_i,
\end{eqnarray}   

  in which $\mathbf{I}$ is the vector of nodal values of $I(r)$ ordered by their global numbering. By solving this set of algebraic equation we can derive $\mathbf{I}$ and $I(r)$ using equation (\ref{sumf}). The resultant light curve of such $I(r)$ will go through all data points near peak ($A_i, i=1..N_1$).
  
 To compare FEM with a simpler numerical inversion technique, we apply PIM (Equation \ref{eq:PIM} ) to the magnification-LD equation (Equation \ref{eq:Ft}) as well. If we choose the result to be piecewise constant we get: 
 
  \begin{eqnarray}\label{eq:AIstd}
\mathbf{A}=\mathbf{D}.\mathbf{I},\;\; D_{ij}=\int_{r_j}^{r_{j+1}} r\,A(r,\ell_i)dr
\end{eqnarray}  
 
 As in FEM we use Runge-Kutta adaptive step size method to derive $D_{ij}$s up to precision of $\mathcal{O}(10^{-7})$.

  \subsection{Simulation Details: Classical FEM }
 In this section, we simulate microlensing light curves with the finite-size effect to study the quality of recovered LD profile ($I_{FEM}$) by FEM. We also compare it with the results from the PIM (i.e. Equation \ref{eq:AIstd}). In this simulation, we examine the effect of different parameters i.e. minimum impact parameter, data cadence and error bars in the light curve data on the quality of the recovered LD profile.We simulate the light curves with the parameters of ($p,t_\ast,t_0,\rho_{\ast}$) with taking a Linear Limb Darkening (LLD) profile and normalize it to the unit total flux (NLLD):\begin{equation}I(r)=\frac{1}{\pi}(1-\Gamma(1-\frac{3}{2}\sqrt{1-r^2}) ),\label{eq:Ir}\end{equation} Where $\Gamma$ is the linear Limb Darkening Coefficient (LDC) and it depends on surface gravity ($\log g$), effective temperature ($T_{eff}$) and metallicity ($Z$) of a star. The LLD profiles that is more commonly used is in the following form of  \begin{equation}\label{lld}I(r)=I(0)(1-u_{LLD}(1-\sqrt{1-r^2}), \end{equation}There is a straightforward relation between $\Gamma$ and $u_{LLD}$ that is: $u_{LLD}={3\Gamma}/{(2+\Gamma)}$.Using stellar atmospheric models, one can derive LDCs for stars with different $\log g, T_{eff}, Z$ \citep{Claret00}.

We use equation (\ref{eq:Ft}) to calculate $A_i$ in $N_1$ different moments of $t_i$ starting from $t^{-}_{transit}$ to $t^{+}_{transit}$, the time when the lens enters and leaves the source disk at $t^{\pm}_{transit}=t_0\pm t_{\ast}\,\sqrt{1-p^2}$. 
We do not use outside the range of $t^{-}_{transit}<t<t^{+}_{transit}$ in our analysis as the LD-effect on the light curve is negligible.

For simplicity, we choose uniform cadence of $t_i$ ($t_{i+1}-t_i=\Delta t $) in this simulation. For each $t_i$ the projected angular distance between the lens and the centre of source disk is $l_i=\sqrt{p^2+\frac{(t_i-t_0)^2}{t_{\ast}^2}}$, where $i=1..N_1$; this leads to producing elements in the $\ell-$domain with different sizes, smaller near centre ($\ell=p$) and larger near edge ($\ell=1$). Then we consider an error bar  of $\sigma_i$ for each point of the light curve  and the magnification from the theoretical light curve shifted by a Gaussian distribution with the width of $\sigma_i$, where $\sigma_i$s results from the uncorrelated magnification error bars of the light curves observed by OGLE and MOA \citep{Ch}.
 
 In the next step we discretize the source of the microlensing event by dividing the stellar disk into $N_2$ annuli ($r_j$ where $j=1..N_2$) and choose annuli such that they cover the whole stellar disk. We note that in the FEM, $N_2$ might be larger than $N_1$ (i.e. $N_2 \geq N_1$).  
Here, in the application of the FEM method, we correspond a map between each data point in the magnification space and the source space where for each element in this space we have 
at least one corresponding data. We have tested that having an annuli-element in the source space with no corresponding data in the magnification space results in a large numerical error.

Moreover, we add a constrain from the physics of the LD to the set of algebraic equation. For the intensity of star at the centre where we call it $I(r_1)$ according to our convention, the radial derivate along $r$-coordinate is zero. This means that $I(r_2) \simeq I(r_1)$  where $r_2$ is the second nodal value in the $r$-space.  We use the convention of  $I(r_i) = I_i$ as we introduced in FEM formalism. 
 Using  $I_2 - I_1 = 0$ constrain and equation (\ref{eq:FtFEM}), we have  
 $N_1+1$ equations with $N_2$ unknowns intensities for each annuli (i.e. $I_1, I_2, ..., I_{N_2}$). If we take $N_2=N_1+1$, then we get a unique solution with using a linear algebraic equations solver. Here we use LU\footnote{Lower triangle, Upper triangle. }-decomposition technique \citep{Pr}.

\begin{figure}
\centering
\begin{subfigure}[h]{0.49\textwidth}
	\centering
	%\caption{The simulated light curve }
        \includegraphics[width=\textwidth]{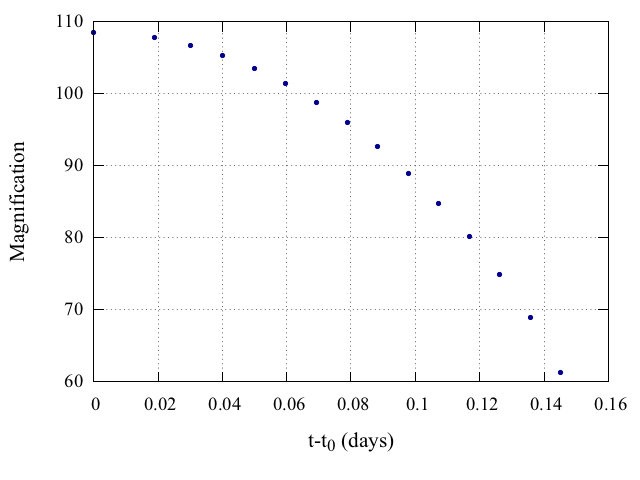}
        \label{fig:A_exact}
\end{subfigure}
\begin{subfigure}[h]{0.49\textwidth}
	\centering
	 % \caption{Results of FEM }
	  %\label{fig:I_A_exact}
        \includegraphics[width=\textwidth]{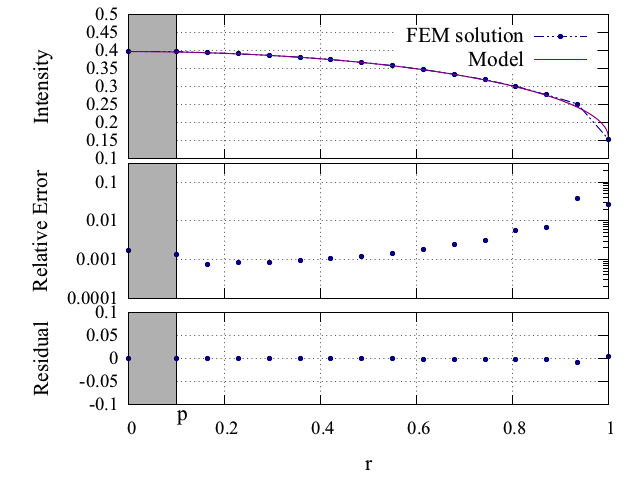}
      \end{subfigure}
\begin{subfigure}[h]{0.49\textwidth}
        \centering
         % \caption{Results of PIM}  
        % \label{fig:I_A_tophat}
        \includegraphics[width=\textwidth]{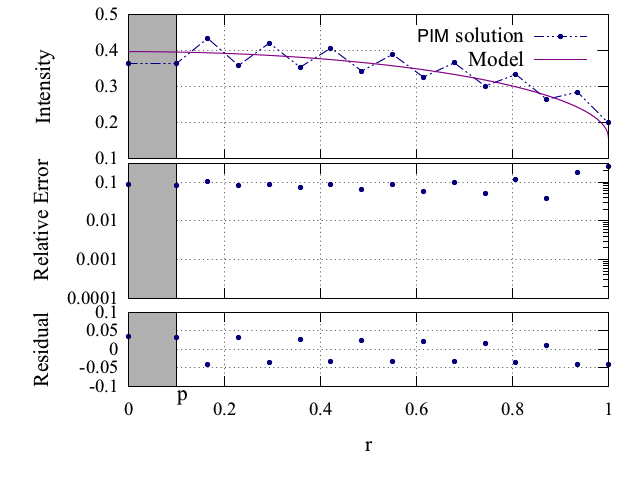}
\end{subfigure}
\caption{The upper panel : The simulated light curve. The middle panel : The recovered LD profile, its relative error and residual compare to input profile by FEM. The lower panel: The same as middle panel for PIM.}
\label{fig:I_of_exactA}
\end{figure}

Now we start with simulation of light curve to examine the FEM. For the first step, we simulate data points of the light curves without taking into account the error bars (i.e. $A_i=A(l_i),\, \sigma_i=0$). The result for the reconstructed LD profile is limited by the errors of the numerical method due to discretisation and the roundoff errors. We take the following set of parameters for our numerical experiment ($p=0.1,\;t_*=0.1458$ days,  $t_0=0,\;\rho_*=0.02,\; u_{LLD}=0.6$) and the cadence of $\Delta t= (t_{transit}-t_0)/(N_1-1), \, N_1=15$. Here the theoretical light curve for this event is shown in Figure (\ref{fig:I_of_exactA}) and the results from reconstructing the intensity of the source are shown in Figure (\ref{fig:I_of_exactA}). We can see that the residuals are larger near the limb as the intensity-derivative of star near the limb is larger than the central part of the star and in order to improve the results, we need more sampling near the limb area. In order to compare the FEM with that of PIM, we plot the result in Figure (\ref{fig:I_of_exactA}). The relative errors and residuals from this method are two order of magnitude higher than the case of FEM.

\begin{figure}
\centering

\begin{subfigure}[h]{0.49\textwidth}
  \centering
  %\subcaption{The simulated light curve data and its error bars ($\frac{\sigma_A}{A}$).}
  \includegraphics[width=90mm]{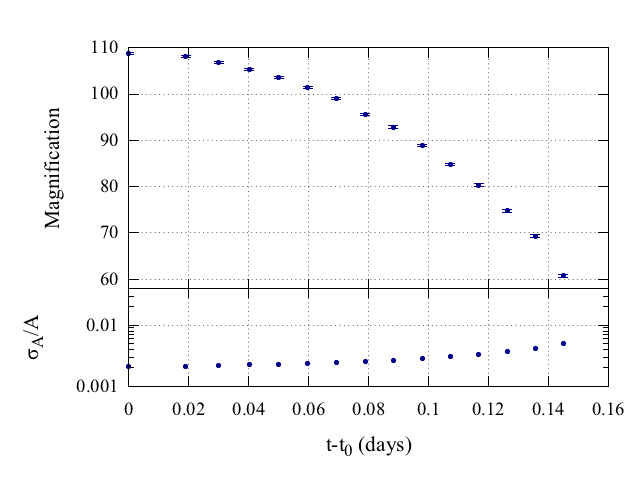}
   % \label{fig:A_dm}
\end{subfigure}

\begin{subfigure}[h]{0.49\textwidth}
   \centering
 % \subcaption{Recovered LD profile ($I_{FEM}$), and its  residual compare to input profile. The purple sade area is the $1-  \sigma$ of the recovered LD profile.} 
   \includegraphics[width=90mm]{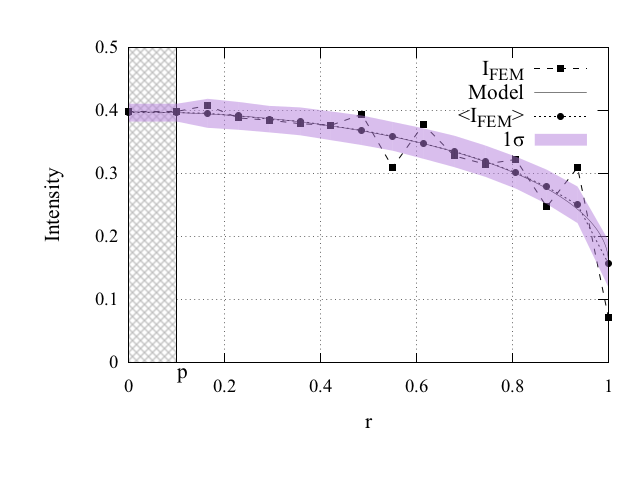}
    % \label{fig:I_A_dm}
 \end{subfigure}
 \caption{ The upper panel: The simulated light curve data and corresponding relative error bars (${\sigma_A}/{A}$) for each data point. The lower panel: Mean value of recovered LD profiles ($<I_{FEM}>$) of $1000$ realization for the light curve in the upper panel (Dotted line).  The shaded area is the standard deviation of  the $1000$ recovered LD profiles. The dashed line is one of recovered profiles. 
 }  
 \label{fig:I_of_A_dm}
\end{figure}

In the next step we study the effect of error bars on the reconstructed intensity from the FEM. We simulate a microlensing event with the parameters given in the first part of this section and the uncorrelated error bars from real MOA and OGLE observations. The average value of the error bars in terms of the magnitude is around  $0.005$.
Then we use the Monte Carlo simulation and produce $1000$ realization from the same event where each event is different than the other in terms of the measure value of the magnification which is given according to the Gaussian error-bar. Fig.\ref{fig:I_of_A_dm} represents a light curve and associated error bar for each data point. We take the mean value of 1000 solutions as the nodal values and associate the error bars of $I_{FEM}$ from the variance $$\sigma_{I_{FEM}}(r_i)=\sqrt{<I_{FEM}(r_i)^2>-<I_{FEM}(r_i)>^2}.$$ 

The results is shown in Fig.\ref{fig:I_of_A_dm}. The average profile of the reconstructed profile (i.e. $<I_{FEM}>$) is close to the simulation's input profile. However, if we take randomly a reconstructed profile from FEM, there are dispersions around some of the points from the input LD profile.  However to overcome this issue we can use regularization techniques \citep{CB86} where we will describe it in the next section.

\begin{figure}
\centering
\includegraphics[width=90mm]{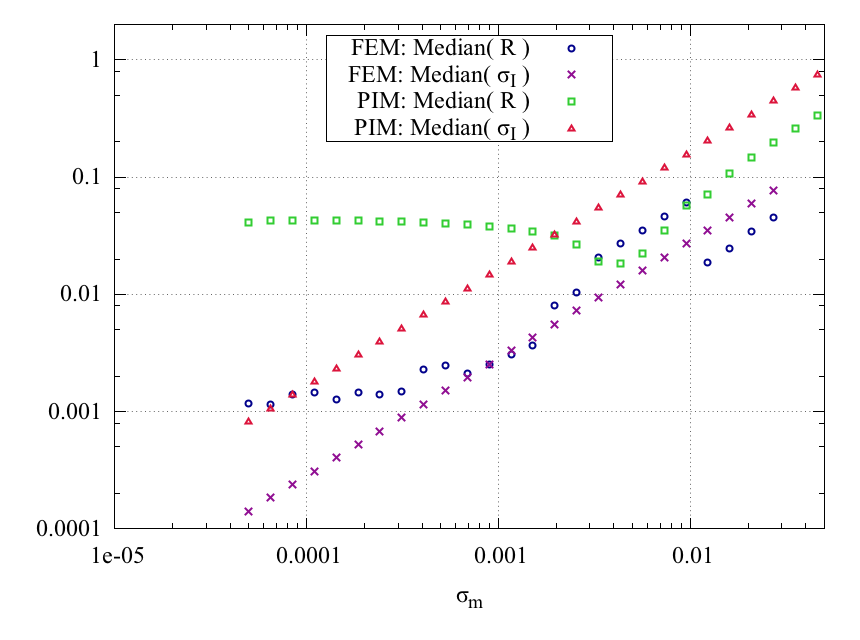}
\caption{The effect of error bars $\sigma_m$ (in terms of apparent magnitude) in the light curve data on the quality of recovered LD profile  by FEM  and PIM. The x-axis is the $\sigma_m$ averaged over all data points. The y-axis is the median of variance ($\sigma_I$) over all nodal values of recovered LD profiles obtained from $1000$ simulated microlensing light curves and median of absolute residual ($R$) over nodal values of the recovered LD profiles. } 
\label{fig:errA_effect}
\end{figure}

We examine the effect of photometric precision of data on recovered intensity profile, By Monte Carlo simulation, we produce $1000$ realisation for a set of events  ($p=0.1,\;t_*=0.1458$ days,  $t_0=0,\;\rho_*=0.02,\; u_{LLD}=0.6, N_1=10$)) with different set of $\sigma_m$. For each set of light curve we calculate $\sigma_{I_{FEM}(r_i)}$ and in order to have an overall dispersion around the FEM result, we calculate the median of  $\sigma_{I_{FEM}(r_i)}$, denoted by median($\sigma_I$) for whole of stellar-disk. This parameter provides a relation between the dispersion of the final result of  FEM to the photometric accuracy.  The other relevant parameter for the quality of the FEM is the median of absolute residuals $|I_{model}(r_i)-I_{FEM}(r_i)|$, also denoted by median($R$).
 The results are shown in Figure (\ref{fig:errA_effect}) where by reducing the photometric errors, the precision of LD data becomes better and converging to the model however there is a limit which might be depend on the sampling rate, data coverage and numerical errors. The same procedure is done for PIM. The result is shown in Figure (\ref{fig:errA_effect}), we see that median($\sigma_I$) for FEM is one order of magnitude smaller compare to PIM and we have the same situation for median($R$) comparing FEM with the PIM.
  
\begin{figure}
\centering
\includegraphics[width=90mm]{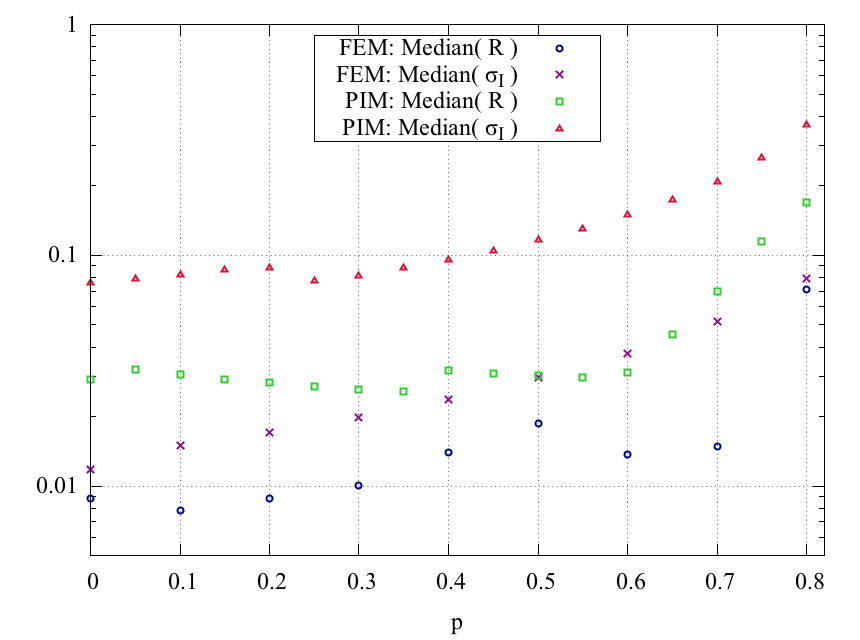}
\caption{ The effect of impact parameter (x-axis) on quality of recovered LD profile (median of $\sigma_I$ and absolute residual $R$, y-axis) from 10 data points using FEM and PIM. }
\label{fig:p_effect}
\end{figure}

Next we study the effect of impact parameter on the quality of the reconstructed LD profile. We fix data cadence ($N_1=10$) and $\sigma_{m,i}\simeq0.005$ while changing impact parameter. Results are shown in Figure (\ref{fig:p_effect}) where the reconstructed function of LD is in favor of the small impact parameters.

\begin{figure}
\centering
\includegraphics[width=90mm]{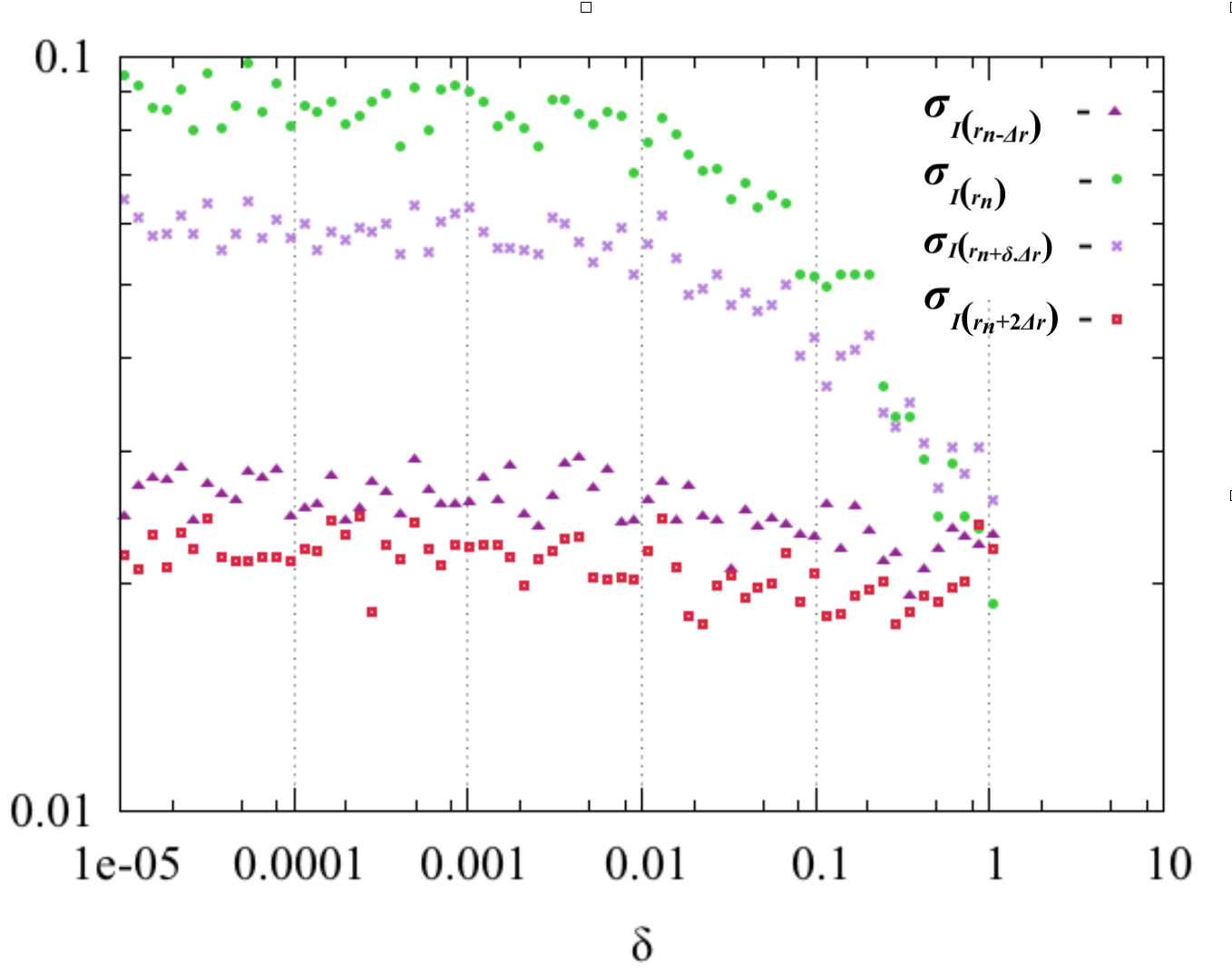}
\caption{Variances of the FEM LD profile in 4 adjacent nodes (y-axis), induced by data with different degree of uniformity around nth data point ($\delta=(t_n-t_{n-1})/(t_{n+1}-t_n)$, x-axis). $\delta=1$ corresponds to a uniform data set. }
\label{fig:dt_min_effect}
\end{figure}

The other important parameter is the uniformity of the data in the light curve. In our simulations we found that two neighboring data points being closer compared to the average time steps of the data set results in a larger error of the corresponding LD nodal value.
To study this effect,  we produce data with uniform cadence ($t_{i+1}-t_i=\Delta t$) then we put two neighboring  points close together ($t_n=t_{n-1}+\delta \Delta t , \, 0 < \delta \le 1$) while other points remain in the same place.
Then for each $\delta=(t_n-t_{n-1})/(t_{n+1}-t_n)$ we produce 1000 light curves and we recover LD nodal values and calculate the variance of $I_{FEM}$ in $r_{n-1}=\ell(t_{n-2})$,  $r_n=\ell(t_{n-1})$, $r_{n+1}=\ell(t_{n})$, $r_{n+2}=\ell(t_{n+1})$ to compare variances of nodal values close to this defect. 
 Results are shown in Figure (\ref{fig:dt_min_effect}) where the variances of $I(r_{n+2})$ and $I(r_{n-1})$ remain similar to the uniform sampling rate ($\delta=1$) however the variances at the denser sampling rate becomes higher (i.e. $r_{n+1}$ and $r_n$).
Most real light curves are not uniform, in these cases we suggest to select several almost uniform $(\delta \gtrsim 0.5)$ subsets of data and recover intensity profile for each subset ($\mathbf{I}_{i}, i=1..N_{subset}$) and then take the average of all to find the final intensity profile.

\begin{figure}
\centering
	\includegraphics[width=90mm]{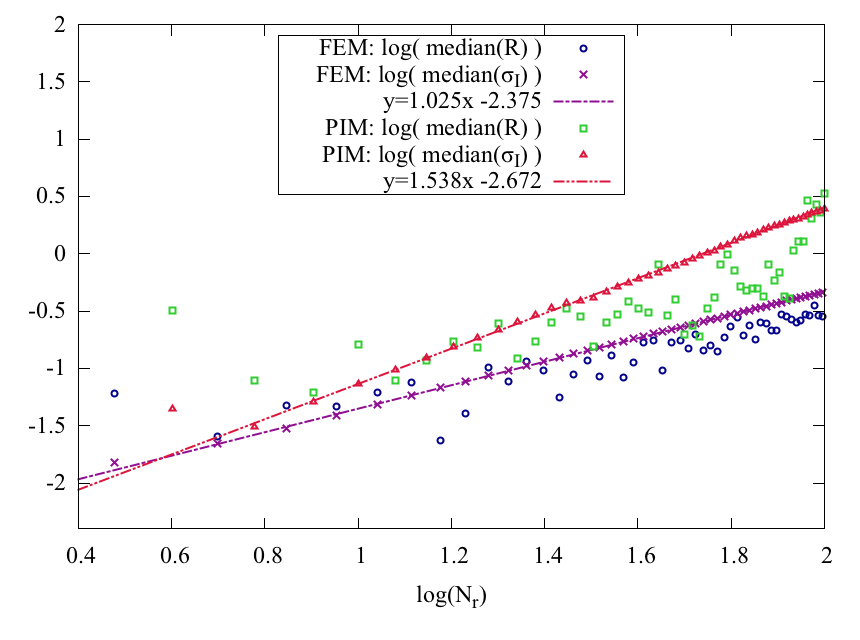}
\caption{ The effect of number of data in the light curve (x-axis) on quality of recovered LD profile by FEM and PIM. The y-axis is the median of variance ($\sigma_I$) over all nodal values of recovered LD profiles obtained from $1000$ simulated microlensing light curves and median of absolute residual ($R$) over nodal values of one of the recovered LD profiles. Fitting the FEM and PIM with a straight line, shows that errors in FEM and PIM depends on the number of data points with the exponent of  approximately $1$ and  $1.5$, respectively.} 
\label{fig:N_effect}	
\end{figure}

To study the stability of solution in FEM in terms of the number of data points, we adapt a constant  ${\sigma_i}$ and let $N_1$ to change from $4$ to $90$. We take a uniform cadence of $\Delta t=(t_{transit}-t_0)/{(N_1-1)}$ to generate light curves. 
The results are shown in Figure (\ref{fig:N_effect}) where increasing the number of data points results in reconstruction of poor LD profile. This effect results from numerical errors and this problem is well known in inverse problems \citep{CB86}. A larger data sets leads to larger  Global  Stiffness Matrix (GSM) and there is more chance that its rows and columns become nearly linear dependent and GSM becomes near singular.  
  
   On the other hand for smaller $N_1$, the average of variance decreases but absolute residual increases, which results in inaccurate intensity profiles. For the case of uniform data cadence and  specified set of primary parameters, the optimum number of data during the transit is obtained to be $N_{1,optimum}=10$. This number depends on the uniformity of data points, error bars and coverage during the transit.  We can stabilize the FEM solutions using regularization technique and we will discuss it in the next section.

In this section we find out that for reducing artificial dispersion of the recovered intensity profiles and to overcome instability problem we need to use regularisation technique. Additionally to reduce the dispersion that results from the denser parts of data points in the light curve  ($\delta \ll 0.5$) we propose to select data subsets that are almost uniform $(\delta \gtrsim 0.5)$ and recover LD profile for each subset and take average over them to find the final LD profile.
We find out that light curves with smaller impact parameter and higher quality photometry give more reliable intensity profiles.

  \subsection{Simulation Details: Regularized FEM }
In the previous section we studied the effects of photometric error bars and increasing the number of data points on the quality of recovered LD profiles. We found out that photometric error bars resulted in artificial dispersions in the recovered intensity profiles and increasing the number of data points increase the dispersions. We can overcome both of these problems using regularisation techniques. The general idea is that instead of solving the equation (\ref{eq:FtFEM}) that returns answers with minimum residual $R^2=\sum_{i=1}^{i=N_1}(A_{data,i}-A_{I_{FEM},i})^2$, we minimize an objective function that combines  $R^2$ with other physical assumptions such as minimizing dispersion of the intensity profile in the nodes (in another word, minimizing the norm of the second derivative of the solution). The objective function is defined as follows \citep{CB86}:
\begin{eqnarray} 
\label{32}
\mathcal{O}=||\mathbf{M}.\mathbf{I}_\lambda-\mathbf{A}||^2 + \lambda \sum_{i=2}^{N_2-1}  (\frac{\mathbf{d}^2 I_{\lambda}}{\mathbf{d}r^2}|_{r=r_i})^2, 
\end{eqnarray} 
in which $\lambda$ is the smoothing parameter.
To minimize this objective function we write this equation in the matrix form and differentiate it with respect to $I_\lambda$. The result is as follows:
\begin{equation}\label{eq:smoothI}
(\mathbf{M}^T\mathbf{M}+\lambda \mathbf{Q})\mathbf{I}_\lambda= \mathbf{M}^T\mathbf{A},
\end{equation}
where $\mathbf{Q}$ is the smoothing matrix. The answer with $\lambda=0$ is the classical solution of Equation (\ref{eq:FtFEM}), as shown in Figure (\ref{fig:I_of_A_dm}) and the larger $\lambda$ yields smoother solutions. At large $\lambda$s the second term is more prominent and the resultant profiles approach to a straight line since the second derivative of a straight line is zero. So further smoothing of the solution provides a straight line for the intensity profile which is not our desire. We change the variable of the intensity profile from $r$ to  $\mu=\sqrt{1-r^2}$ in the second term of equation (\ref{32}) 
which results in the extension of limb area and it is suitable for the recovering the limb of the source star. The second finite derivative of intensity profile in terms of the new variable, $\mu$ is 
$$ \frac{\mathbf{d}^2 I_{\lambda}}{\mathbf{d}\mu^2}|_{\mu=\mu(r_i)}=$$
$$(\frac{2}{\Delta_i \Delta_{i-1}})^2 \lbrack \frac{\Delta_i}{\Delta_{i-1}+\Delta_i} I_{\lambda,i-1}-I_{\lambda,i} + \frac{\Delta_{i-1}}{\Delta_{i-1}+\Delta_i}I_{\lambda,i+1} \rbrack ,$$
in which $\Delta_i=\mu(r_{i+1})-\mu(r_i)$. We note that $\Delta_i$s are not equal. According to equation (\ref{eq:smoothI}) there exists a solution of $\mathbf{ I_\lambda }$ for each $\lambda$. To recover LD profile for each light curve we need to find the optimum value of $\lambda$ using simulations. We generate simulated light curve data with a similar cadence and photometric error bars as in the real data. Also we implement our known limb darkening profile ($I_{input}$) in the light curve.

In the real light curves, we have non-uniform data cadence and some of data points are very close to each other ($\delta < 0.001$). From our investigation in the previous section, to avoid errors from the non-uniform sampling in the recovered profiles, we need to generate almost uniform data subsets. We recover LD profiles for each of the subsets and take the average over all recovered profiles. 
In order to select almost uniform subsets of data we produce histogram of data with number of $N_{hist}$ bins for each light curve. Then, we randomly select one data from each of the bins of the histogram to produce an almost uniform ($ median(\delta) \gtrsim 0.5 $) subset of data. We generate an ensemble of data subsets to use all the initial data in the observed light curve. If the original light curve satisty the condition $ median(\delta) \gtrsim 0.5 $ we use all data at once and skip bininng procedure. 
The final result is the average of all the profiles (i.e. $I_{FEM,\lambda}=1/N_s\sum_{i=1}^{N_s} \mathbf{ I_{\lambda, i} } )$. 

 Here, we set $N_{hist}=100$, also since $\lambda$ is a free parameter, we let this parameter change 
with a power-law function as $\lambda_i=10^{i}$ (instead of a linear function). We choose the parameter of $i$ in the range of $i\in[-5,3]$ with $160$ steps. 
For each $\lambda_i$  we calculate dispersion ($D_{\lambda_i}$) and squared weighted 
residual ($R_{\lambda_i}$) of the corresponding $I_{FEM,\lambda_i}$:
\begin{equation}
D_{\lambda_i}=\sum_{j=1}^{N_2}\sum_{k=1}^{N_2}I_{FEM,\lambda_i,j}Q_{jk}I_{FEM,\lambda_i,k},
\label{eq:dsprn}
\end{equation}
in which $I_{FEM,\lambda_i,j}$ is the $j$th node of $I_{FEM,\lambda_i}$ and 
\begin{equation}
R_{\lambda_i}=\frac{1}{N_1}\sum_{j=1}^{N_1}(\frac{A(I_{FEM,\lambda_i})_j - A_j}{\sigma_j})^2,
\label{eq:wR2}
\end{equation}
 in which $A(I_{FEM,\lambda_i})_j$ is the recovered amplification with the parameter of $\lambda_i$. 
 A profile ($I_{FEM,\lambda_i})$ with $R_{\lambda_i} \simeq1$ would provide the reasonable recovered profile.  On the other hand, a profile with $R_{\lambda_i}\eqsim0$ is an oscillatory solution and it generates a light curve that goes through all data points.
 
To find the optimum value of $\lambda$ we use a combined condition using both $D_{\lambda_i} $ and  $R_{\lambda_i}$ functions. First we find minimum dispersion $min(D_{\lambda_i})$ and check $R_{\lambda_i}$ to satisfy the following condition of 
\begin{equation}
 min|R_{\lambda_i}-1|~  \backepsilon ~ [min(D_{\lambda_i})<D_{\lambda_i}<\alpha\times min(D_{\lambda_i})],
 \label{eq:condition}
\end{equation} 
 where $\alpha>1$ and this parameter corresponds to a neighborhood around the minimum value of $min(D_{\lambda_i})$. In order to specify the $\alpha$ value, we use a test with simulating light curves where data points have the same cadence and error bars as in the real light curve while functions of  $I_{input}$s are different. We call these as the control simulations and take one of the input profiles to be a square root function: 
\begin{equation}
\label{sqrt}
I_{Sqrt}(r)=f_1\left(1-c(1-\sqrt{1-r^2})-d(1-\sqrt[4]{1-r^2})\right)
\end{equation}
 and the other one to be
 \begin{equation}
 \label{tanh}
 I_{tanh}(r)=f_2\left(1-\tanh{( 4(1-\sqrt{1-r^2})-2.5)}\right),
 \end{equation}
where $f_1$ and $f_2$ are chosen in such a way to normalize the overall flux of source star to be one. 

\begin{figure}
 \centering
 \includegraphics[width=90mm]{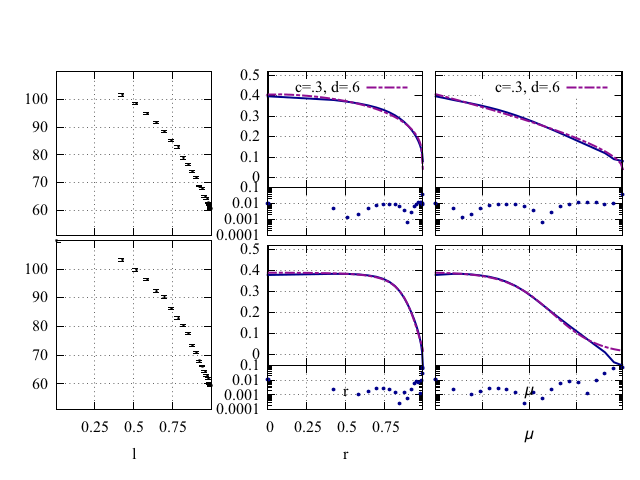}
 \caption{Reconstructed LD profiles from control simulations, using regularized FEM with the condition from equation (\ref{eq:condition}) and $\alpha=4.5$. The left panels shows the light curves (Amplifications in terms of $\ell$) with 20 data points and ${\sigma_A}/{A}=0.005$, the right panels shows the reconstructed LD profiles (solid lines) together with the input profiles (dashed lines) both in terms of $r$ and $\mu=\sqrt{1-r^2}$. The residual between the the recovered and input profiles are also plotted with blue points. We skip the binning procedure and used all the data points without making subset of data points.}
 \label{fig:20data}
 \end{figure}

\begin{figure}
 \centering
 \includegraphics[width=90mm]{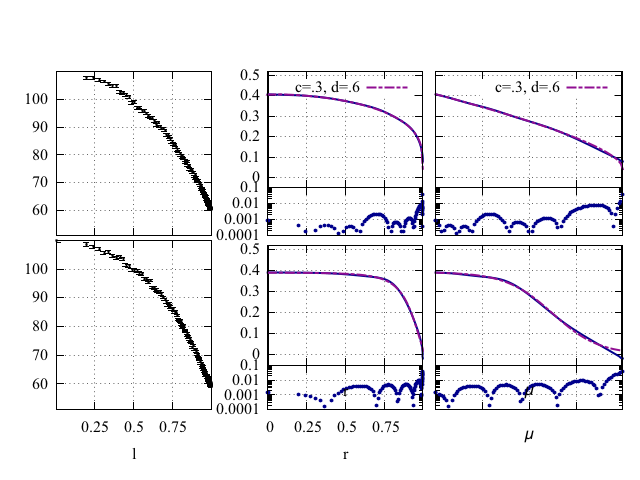}
 \caption{Reconstructed LD profiles from control simulations, using regularized FEM with the condition from equation (\ref{eq:condition}) and $\alpha=2.5$. The left panels shows the light curves (Amplifications in terms of $\ell$) with 100 data points and ${\sigma_A}/{A}=0.005$, the right panels shows the reconstructed LD profiles (solid lines) together with the input profiles (dashed lines) both in terms of $r$ and $\mu=\sqrt{1-r^2}$. The residual between the the recovered and input profiles are also plotted with blue points. We skip the binning procedure and used all the data points without making subset of data points.}
 \label{fig:100data}
 \end{figure}

These two function behave differently in $\mu$-space. In the first function from equation (\ref{sqrt}), $d^2 I(\mu)/d\mu^2<0$ while the second function in equation (\ref{tanh}) has reflection point with $d^2 I(\mu)/d\mu^2=0$. From the first profile in equation (\ref{sqrt}) we obtain the best solution and from the second profile we exclude the the recovered profiles with the overly smoothed solutions; such a solution is very close to a straight line in the $\mu$ space regardless of the input profile.  Figures (\ref{fig:20data}) and (\ref{fig:100data}) show the result of this procedure for a microlensing event with two different number of data points in the light curve and a high cadence near limb. We use all data at once and skip the binning procedure. Here is the parameters of the light curve: $\sigma_A/{A}=0.005$ and the lensing parameters as $\rho_\ast =0.02, p=0.005, t_E=7.29 \text{ days}, t_0=0, c=0.3, d=0.6$. 
 
\subsection{Results}

We apply FEM to a sample of microlensing light curves with finite-source effect to obtain the limb darkening of the source stars.  This sample is discovered by survey groups  of MOA and/or OGLE and alerted to the follow-up collaborations of PLANET,  $\mu$FUN, RoboNet, MiNDSTEp. These events have been analyzed by \citet{Ch} where they assumed a linear standard limb darkening function and found parameter of the profile from fitting to the light curves in a specific filter. 
Here, we choose $7$ light curves from six events that have a good data coverage near the peak. We note that the limb darkening is a wavelength dependent phenomenon and data of each observatory with different filters should be analyzed separately. The selected light curves are listed in Table (\ref{tab:LCs}) with the best values of the lensing parameters in \citet{Ch}. Also from the standard $\chi^2$-fitting, we obtain the blending (i.e. $b$) and the baseline of the source stars (i.e. $m_{BL}$).  We used high quality data as in \citet{Ch} with $t<3t_E$, $\sigma_m<0.1$ and $residual<3\sigma_m$.

\begin{figure}
\includegraphics[width=90mm]{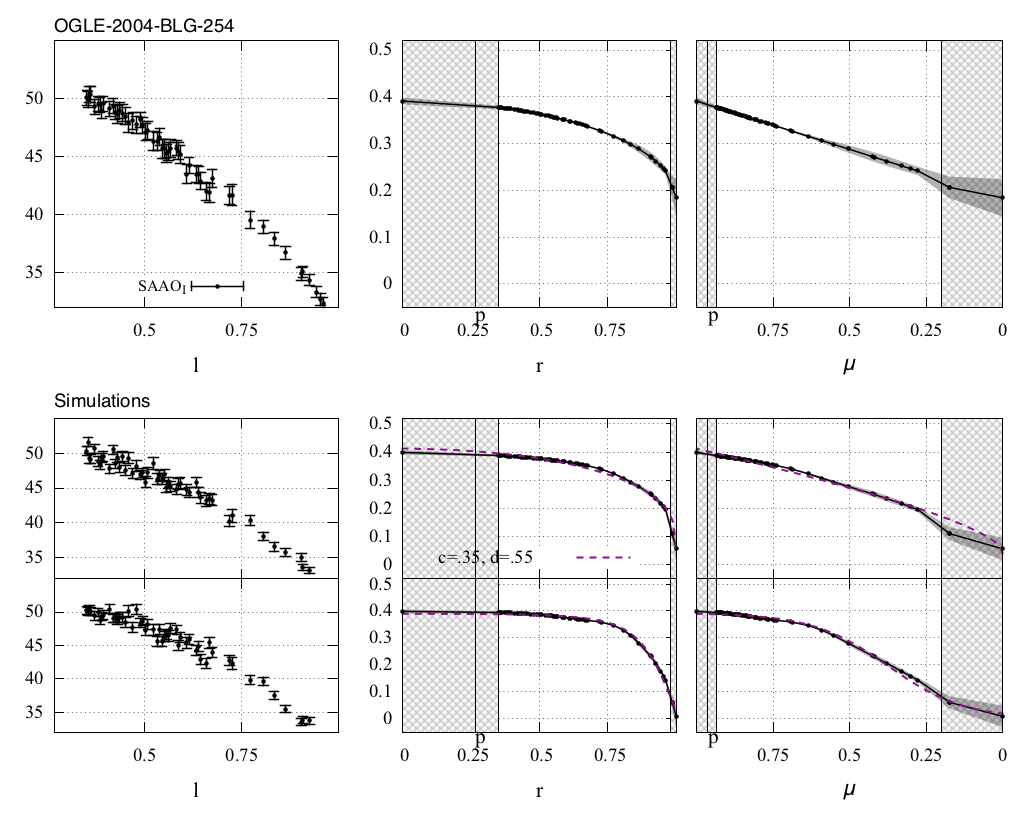}
\caption{The top left side is the light curve of OGLE-2004-BLG-254 SAAO I-band data 
versus angular projected distance ($\ell$). The top right panel is the recovered intensity profiles in $r$ and $\mu$ space with the regularised FEM. The dashed grey corresponds to the region where no data is available and in the white region there is observed data. The lower panels are the control simulations with the profile from equation (\ref{sqrt}) and equation (\ref{tanh}), respectively. The input profiles for simulation are in dashed lines and the reconstructed profiles are in black points connected by the solid line.}
\label{fig:LC1}
\end{figure}

\begin{figure}
\includegraphics[width=90mm]{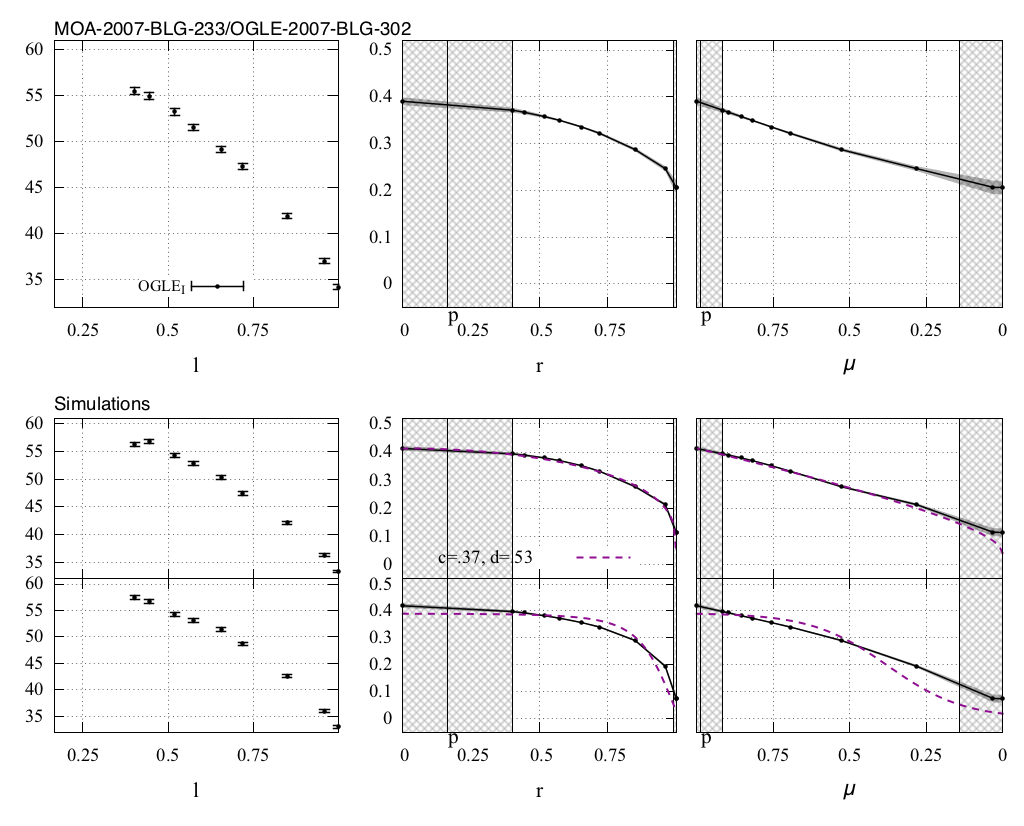}
\caption{The analysis of MOA-2007-BLG-233/OGLE-2007-BLG-302 even. In this figure we use OGLE I-band data. The detailed information about this figure is the same as figure (\ref{fig:LC1}).}
\label{fig:LC2}
\end{figure}

\begin{figure}
\includegraphics[width=90mm]{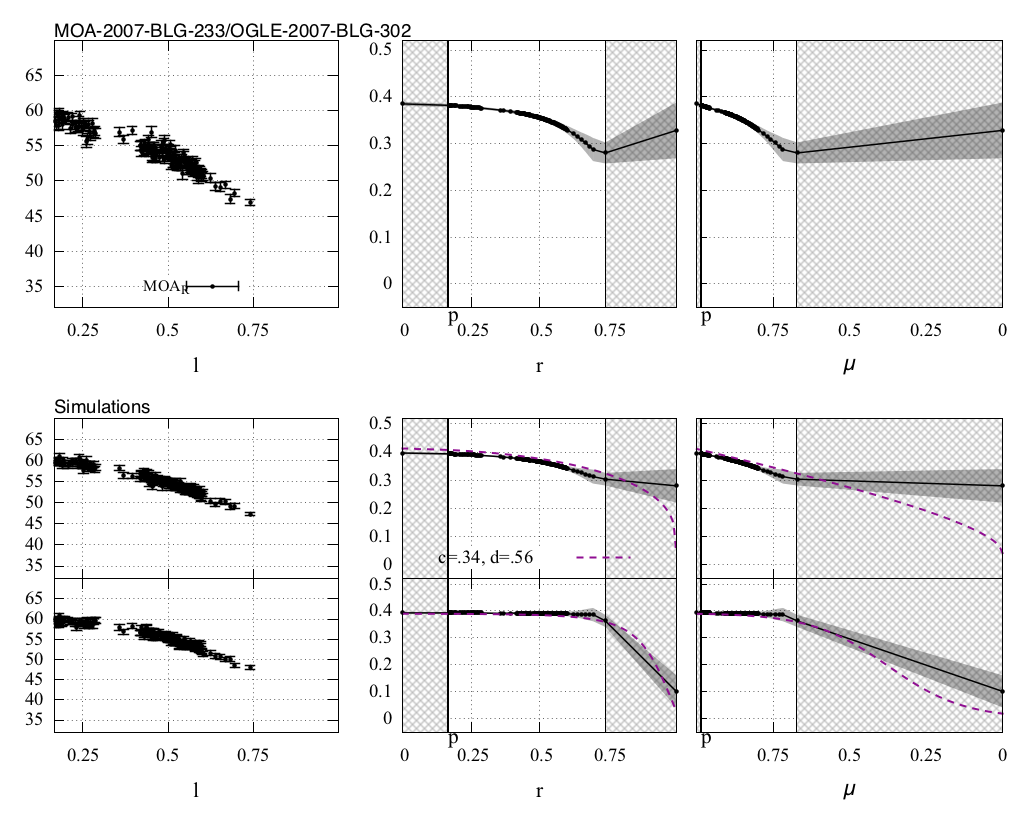}
\caption{The analysis of MOA-2007-BLG-233/OGLE-2007-BLG-302 even. In this figure we use MOA R-band data. The detailed information about this figure is the same as figure (\ref{fig:LC1}).}
\label{fig:LC3}
\end{figure}

\begin{figure}
\includegraphics[width=90mm]{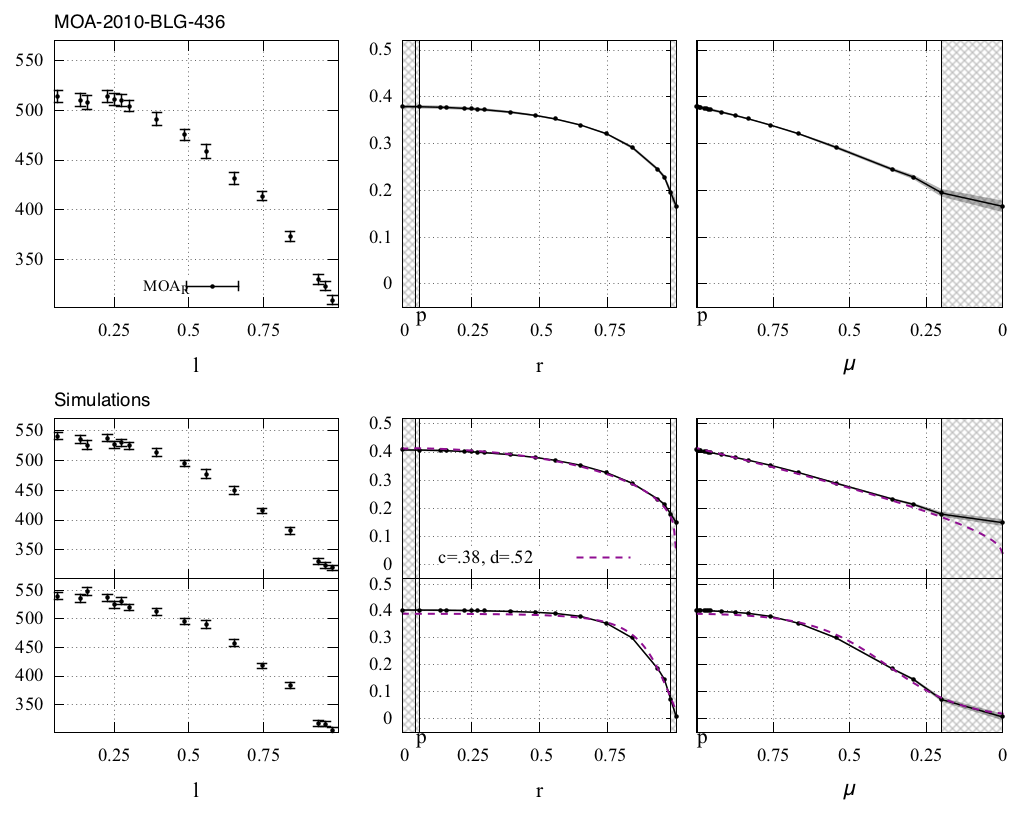}
\caption{The analysis of MOA-2010-BLG-436 in R-band. The detailed information about this figure is the same as figure (\ref{fig:LC1}).}
\label{fig:LC4}
\end{figure}

\begin{figure}
\includegraphics[width=90mm]{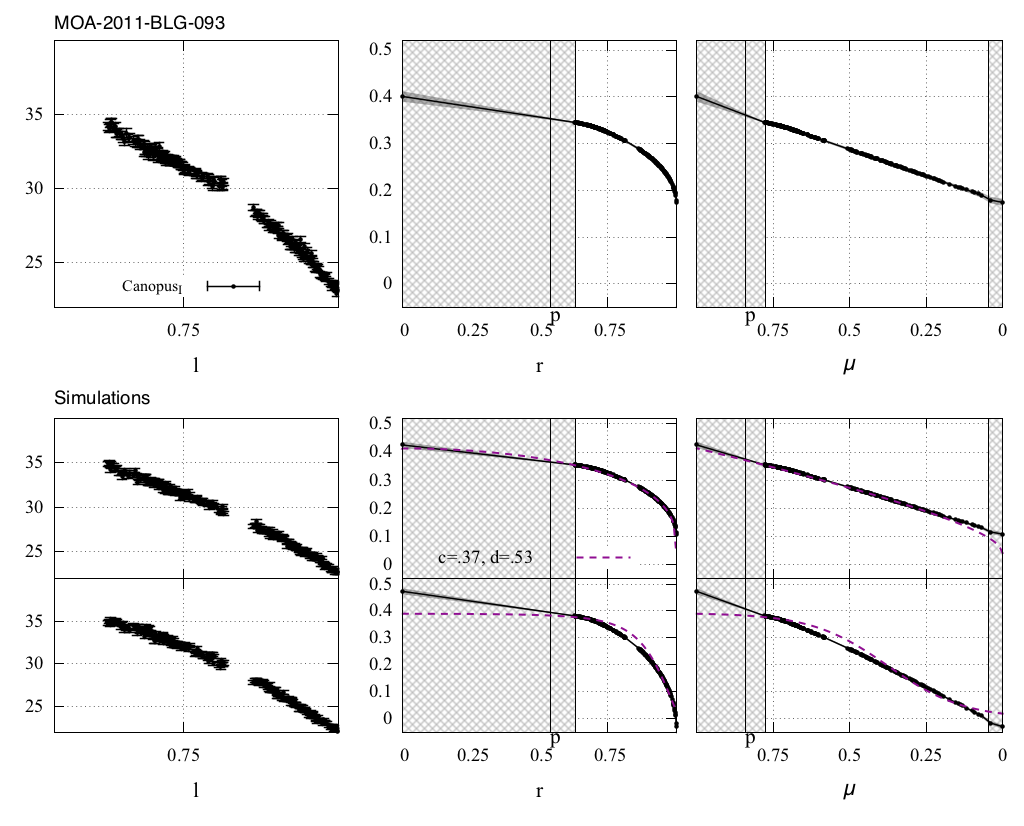}
\caption{The analysis of MOA-2011-BLG-093 from Canopus Hill Observatory's I-passband data. The detailed information about this figure is the same as figure (\ref{fig:LC1}).}
\label{fig:LC5}
\end{figure}

\begin{figure}
\includegraphics[width=90mm]{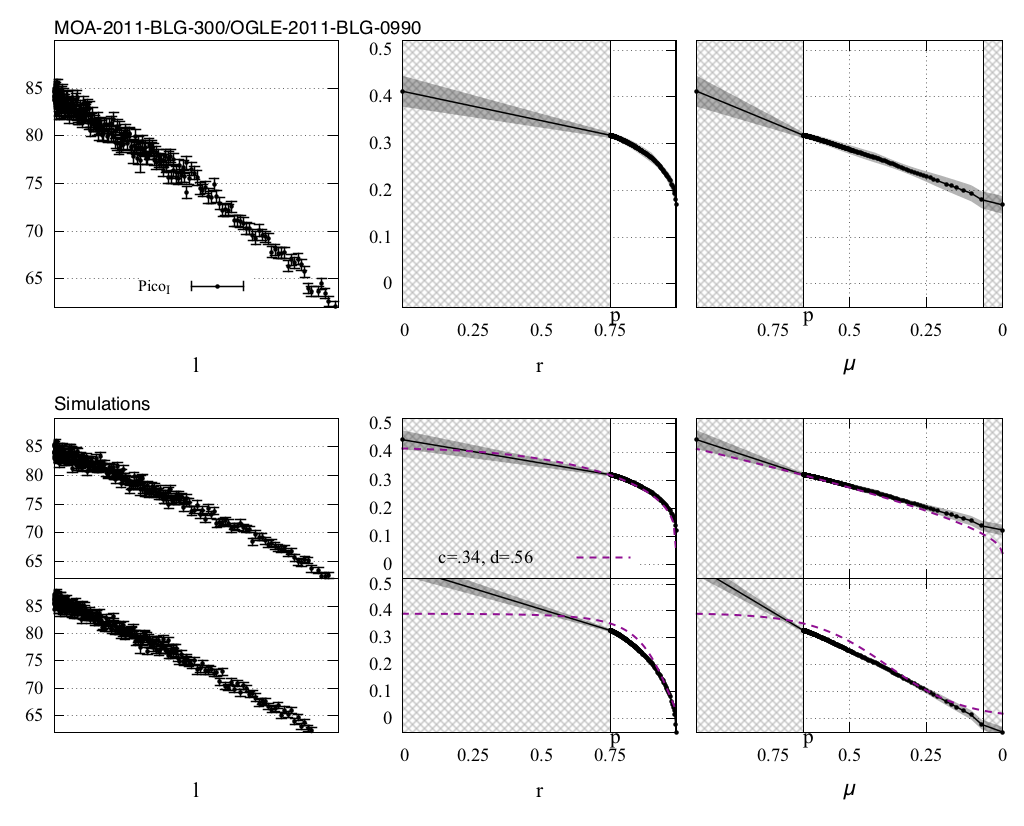}
%\caption{Light curve of MOA-2011-BLG-300/OGLE-2011-BLG-0990, $Pico_I$ observations the control simulations and their recovered profiles. Notations are the same as Figure (\ref{fig:LC1})}
\caption{The analysis of MOA-2011-BLG-300/OGLE-2011-BLG-0990. We use data from Observatorio do Pico dos Dias in I-passband. The detailed information about this figure is the same as figure (\ref{fig:LC1}).}
\label{fig:LC6}
\end{figure}

\begin{figure}
\includegraphics[width=90mm]{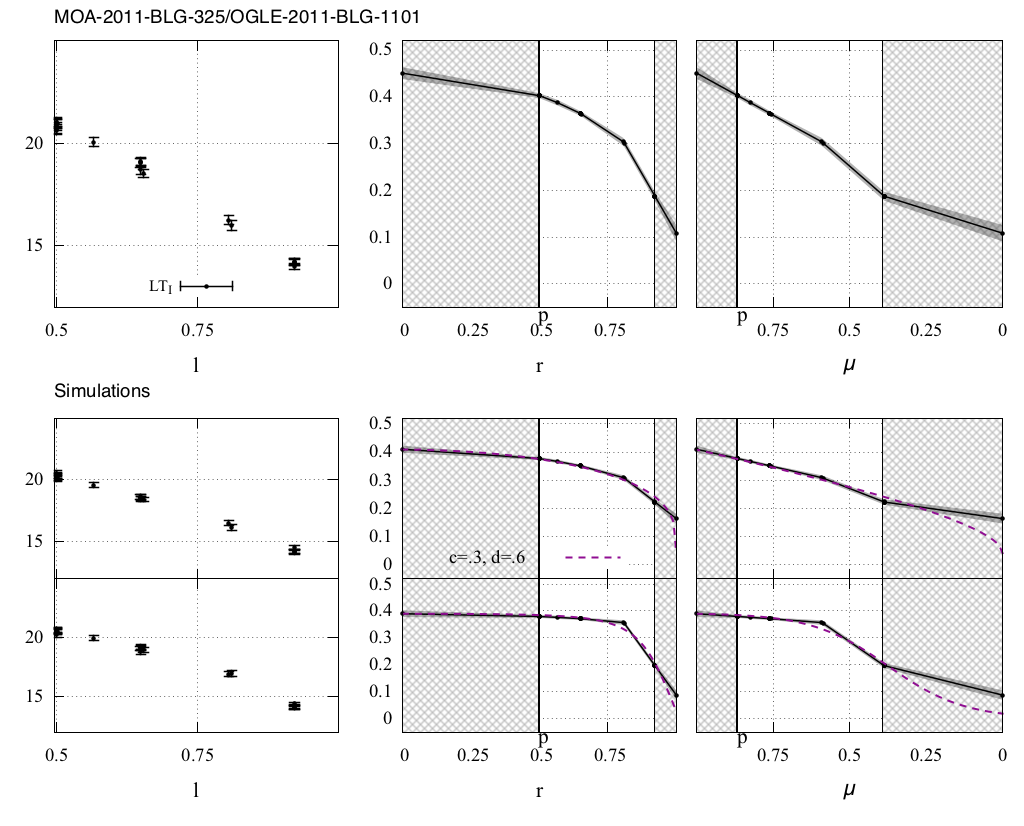}
\caption{The analysis of MOA-201-BLG-325/OGLE-2011-BLG-1101. We use data from Liverpool telescope in I-passband. The detailed information about this figure is the same as figure (\ref{fig:LC1}).}
\label{fig:LC7}
\end{figure}

\begin{table*}
	\centering
	\caption{First column represents the name of microlensing event and the abbreviation name of the observatory that data are taken for our analysis. Here we adapt a single filter light curve. The next four columns are the primary parameters we used in our analysis obtained by \citep{Ch}. Next two columns  are the magnitude of baseline and the blending parameter. The other columns are the average of uncertainties in derivation of $ \ell  \; (\sigma_\ell) $ and number of data with $ \ell <1 $. }
	\label{tab:LCs}
	%\begin{tabular}{ | p{3cm} | p{1cm}| p{1cm} | p{1cm} | p{1cm} | p{1cm}| p{1cm}| p{1cm}|p{1cm}|}
	\begin{tabular}{ | c | c| c | c | c | c| c| c|c|}
		\hline
	         Event   &$t_0$(HJD-2450000)&$u_0$ &  $t_E$ (days) & $\rho_\ast$ &  $m_{BL}$ &b & average of $\sigma_\ell$ & No. of data   \\
	          ($Observatory_{~passband}$ )                 &            &             &           &           &             &           &   	     &  ($\ell <1$)  \\
		\hline
		OGLE-2004-BLG-254   & $3166.823$& $0.0111$       & $12.84$    & $0.0418$       & 16.33 & 1.039  & 0.01   &  56 \\
		     ($SAAO_I$)            & $ \pm0.001 $& $\pm0.0004$&$\pm0.09$ & $\pm0.0004$ &$\pm0.014$ & $\pm0.032$ &   	     &    \\
		\hline
		MOA-2007-BLG-233/   & $4289.269$ & $0.0060$& $15.90$ & $0.0364$ & 16.31 & 1.021   & 0.004 &  9 \\
		OGLE-2007-BLG-302   &      $\pm0.001$      &     $\pm0.0002$        &     $\pm0.05$      &     $\pm0.0001$      &   $\pm0.013$ & $\pm0.033$  &   	     &    \\
		     ($OGLE_I $)            &            &             &           &           &             &           &   	     &    \\
		\hline
		MOA-2007-BLG-233/   & $4289.269$ & $0.0060$&$15.90$ & $0.0364$& 16.31 & 1.021  & 0.004 &  143 \\
		OGLE-2007-BLG-302   &      $\pm0.001$      &     $\pm0.0002$        &     $\pm0.05$      &     $\pm0.0001$      &  $\pm0.013$ & $\pm0.033$  &   	     &    \\
		     ($MOA_R $)            &            &             &           &           &             &           &   	     &    \\
		\hline
		MOA-2010-BLG-436   & $5395.791$& $0.0002$ & $12.78$ & $0.0041$ &16.96 & 0.026 & 0.092 & 16 \\
		     ($MOA_R$)           &      $\pm0.001$      &     $\pm0.0002$        &     $\pm1.08$      &     $\pm0.0003$      &$\pm0.038$ & $\pm0.003$&   	     &    \\
		\hline
		MOA-2011-BLG-093   & $5678.555$& $0.0292$ & $14.97$ & $0.0538$ & 16.7 & 1.8 & 0.005 &  206\\
		       ($Canopus_I$)    &      $\pm0.001$      &     $\pm0.0002$        &     $\pm0.05$      &     $\pm0.0002$      & $\pm0.01$ & $\pm0.052$  &   	     &    \\
		\hline
		MOA-2011-BLG-300/  & $5758.691$ & $0.0151$& $6.70$ & $0.0199$ & 18.49 & 0.983 & 0.023 &   229\\
		OGLE-2011-BLG-0990 &      $\pm0.001$      &     $\pm0.0004$        &     $\pm0.07$      &     $\pm0.0003$      &$\pm 0.05$ & $\pm0.3$ &   	     &    \\
		      ( $Pico_I$)           &            &             &           &           &             &           &   	     &    \\
		\hline
		MOA-2011-BLG-325/   & $5823.574$ & $0.0485$& $29.06$ & $0.0979$ & 15.18 & 1.038 & 0.06 &  19\\	
		OGLE-2011-BLG-1101&      $\pm0.002$      &     $\pm0.0005$        &     $\pm0.11$      &     $\pm0.0006$      & $\pm0.008$ & $\pm0.014$ &   	     &    \\
		     ($LT_I$)            &            &             &           &           &             &           &   	     &    \\
		\hline
	\end{tabular}
\end{table*}
Figures (\ref{fig:LC1}-\ref{fig:LC7}) represents the light curves of  $7$ events in $\ell$-space (see equation (\ref{ell})) together with their control simulations in the left-down panels and corresponding reconstructed limb darkening of the source stars in r-space at the middle panels and in the $\mu$-space at the right panels. In these figures,  the grey region corresponds to the area that there is no observed data. 
In the white region we have data observed by the telescopes. The light grey area around the recovered profile is the standard deviation 
 that is obtained from ten simulated light curves with the cadence and error bars as the real data. The list of events and corresponding analysis are as follows:\\

%\begin{landscape}

\begin{table*}
%\centering
	\centering
	\caption{ The first column represents the name of microlensing event and the abbreviation name of the observatory that data are taken for our analysis. Here we adapt a single filter light curve.
	The second column is the source type and the third column is the best fitted value of LDC from previous studies, the fourth column is the best value of LDC to the regularised FEM solutions
	that are shown in Figures (\ref{fig:LC1}-\ref{fig:LC7}), the fifth column is the $\chi^2/N_{dof}$ from fitting the LLD profile to the recovered profile. The sixth, seventh columns are the best values of square 
root LDCs (see equation \ref{sqrt}) and the eighth column is the corresponding $\chi^2/N_{dof}$. }
	\label{tab:LDCs}
    \begin{tabular}{| c| c| c| c| c| c| c| c| }
		\hline
			  (1)	Event name			    &(2)  Source type         &(3) Best fitted $u_{LLD}$&(4) $ u_{LLD}$ &(5) $\chi^2/N_{dof}$ & (6) $c$  & (7) $d$  & (8) $\chi^2/N_{dof}$  \\
		(Observatory-passband)		   &( log g, $T_{eff})$  & of previous studies    &		      &	 (LLD)	        &          &       & (square-root)                       \\	
				   					  & Choi, et al. 2012   &  				    &				         &			 &                         &                          &                               \\		
		\hline
		OGLE-2004-BLG-254 &  KIII		    & $0.55\pm0.06$      & $0.575$  & $29.51/56$&           $0.262$              &      $0.48$        &           $8.14/55$            \\
				   				&          &  Choi et al. (2012) &	$\pm0.028$	&        &      $\pm0.03$         &      $\pm0.04$       &                       \\	
			(SAAO-I)	 &  (2.0, 4750 K)   & $0.45^{+0.03}_{-0.06}$ &                             	&                                   &                         &                          &                     \\
					   					  &   &  Cassan et al. (2006) &					&                           &                         &                          &                      \\	
		\hline
		MOA-2007-BLG-233/      & GIII                 &$0.53\pm0.04$ & $0.537$   & $16.00/8$&   $0.698$         &       $-0.25$       &         $7.07/7$             \\
		OGLE-2007-BLG-302   &(2.5, 5000 K)   &   Choi et al. (2012) &  $\pm0.03$            &            	&     $\pm0.03$     &      $\pm0.05$      &                      \\
		   ( OGLE-I  )       	     &   		      &                         &                          &              	&                         &                          &                       \\
		\hline
		MOA-2007-BLG-233/   & GIII &$0.56\pm0.02$ & $0.56$   & $98.38/143$ &      $0.662$        &       $-0.25$    &          $78.95/142$           \\
		OGLE-2007-BLG-302 & (3.0, 5500 K)  & Choi et al. (2012)&  $\pm0.05$     & 				&     $\pm0.04$     &     $\pm0.06$     &                   \\
		    (MOA-R )  	     &   		      &                         &                          &        		&                         &                          &                    \\
		\hline
		MOA-2010-BLG-436 & ... &$0.52\pm0.1$ & $0.51$   & $62.38/16$   &   $0.025$     &      $0.783$       &          $24.75/15$           \\
			(MOA-R)     &      &   	Choi et al. (2012)		&   $\pm0.08$            & 			&       $\pm0.03$      &      $\pm0.04$       &                     \\
		\hline
		MOA-2011-BLG-093 & GIII 	    &$0.51\pm0.03$ & $0.50$  & $1423.62/198$ &   $-0.046$       &      $0.876$       &           $409.76/197$           \\
			( Canopus-I)& (2.5, 5000 K) & Choi et al. (2012)     &     $\pm0.03$     &			&    $\pm0.03$  &    $\pm0.04$      &                      \\
		\hline
		MOA-2011-BLG-300/     & ... &$0.56\pm0.04$ & $0.39$   & $35.93/226$  &     $0.228$     &     $0.456$       &          $1.575/225$          \\
		OGLE-2011-BLG-0990&     &     Choi et al. (2012)                    &  $\pm0.1$     & 			&    $\pm0.14$       &   $\pm0.23$       &                      \\
		 (Pico-I)		   &      &   			     &                           &              		&                         &                          &                       \\
		\hline
		MOA-2011-BLG-325/     &KIII& ... & $0.94$   & $33.32/19$ &  $0.160$        &      $1.217$       &        $10.35/18$             \\	
		OGLE-2011-BLG-1101& (2.0, 4250 K)&    &   $\pm0.07$	&                         &   $\pm0.051$   &   $\pm0.08$     & \\
		(LT-I)		            &                      &   			     &        		&                         &                          &                      & \\
		\hline
    \end{tabular}
\end{table*}

%\end{landscape}

\textbf{OGLE-2004-BLG-254:} 
The  OGLE-2004-BLG-254 is a bulge event discovered by the OGLE survey and alerted for the follow-up telescopes of PLANET and $\mu$FUN collaborations. We use SAAO I-band of PLANET collaboration which has good data coverage and better photometric quality in our analysis. 
This event was analysed for the first time by \citet{Ca06} and later by \citet{Ch}.
They determined the source type to be a KIII star, based on its location in the CMD\footnote{Color-Magnitude Diagram}. They both obtained linear LDC (i.e. $u_{LLD}$ in equation.(\ref{lld})) with $\chi^2$ fiting. \citet{Ca06} used \citet{Go94} approximation for a single lens event  with an extended source of uniform intensity. They derived $u_{LLD}=0.45^{+0.03}_{-0.06}$ for the I-band observation of SAAO, the source size $\rho_{\ast}=0.04\pm0.0002$ and lens parameters ($t_0=3166.8194\pm0.0002, u_0=0.0046\pm0.0008, t_E=13.23\pm0.05$). \citet{Ch} also used inverse-ray shooting technique to compute light curve. They derived $u_{LLD}=0.55\pm0.06$ in I-band observation of SAAO, the source size $\rho_{\ast}=0.0418\pm0.0004$ and lens parameters ($t_0=3166.823\pm0.001, u_0=0.0111\pm0.0004, t_E=12.84\pm0.09$). In our analysis we adapt these lens parameters for our light curve. Since we have good data coverage in this light curve, we choose smaller bins around the limb to recover the limb darkening with better accuracy. Figure (\ref{fig:LC1}) represents the recovered profile intensity profile for this event. 

\textbf{MOA-2007-BLG-233/OGLE-2007-BLG-302:}
This event was discovered and alerted by both MOA and OGLE surveys, the follow-up observations carried out by $\mu$FUN, PLANET and MiNDSTEp collaborations. The observational data suitable for our method (as discussed before) are Las Campanas Observatory's I-passband (OGLE-I) and Mt. John Observatory's R-passband (MOA-R). This event was analysed by \citet{Ch}. The source star is a GIII star and using a best fit parametric method they derived $u_{LLD}=0.56\pm0.02$ for the MOA-R data and $u_{LLD}=0.53\pm0.04$ for the OGLE-I data.
The recovered LD profile by regularised FEM is shown in Figure (\ref{fig:LC2}) in OGLE I-band Figure (\ref{fig:LC3}) in MOA R-band. We note that for this event OGLE has almost uniform coverage of the light curve, we skip the binning procedure. 

\textbf{MOA-2010-BLG-436:}
This event is observed by MOA survey in R-band.
It was analysed by \citet{Ch}. The type of the source star was not determined due to bad quality of data in V-band however the LDC was determined to be $u_{LLD}=0.52\pm0.10$ from MOA-R data. For this event we have uniform data coverage and we skip the binning procedure. Figure (\ref{fig:LC4}) shows the LD from the regularised FEM.

%The uncertainly in deriving $\ell$ nodes ($\sigma_\ell$) for this case is larger compare to the other events as is shown in Table (\ref{tab:LCs}) so the recovered intensity profile may not be reliable. 

\textbf{MOA-2011-BLG-093:}
This event was discovered and alerted by both MOA and OGLE surveys. The follow-up observations were carried out by $\mu$FUN, PLANET, RoboNet  and MiNDSTEp collaborations. The observational data suitable for our method is Canopus Hill Observatory's I-passband (Canopus-I).
This event was analysed by \citet{Ch} and they determined the source star as GIII and the best LDC as $u_{LLD}=0.51\pm0.03$. We had a good data coverage for this event around the limb and we choose smaller binning for this part. Figure (\ref{fig:LC5}) shows the LD from the regularised FEM.
%Figure (\ref{fig:resultset-Data}) compares  \citet{Ch} results with that of ours from FEM. 

\textbf{MOA-2011-BLG-300/OGLE-2011-BLG-0990:}
This event was discovered and alerted by both MOA and OGLE surveys. The follow-up observations carried out by $\mu$FUN, PLANET collaborations. The observational data suitable for our method is taken from Observatorio do Pico dos Dias, I-passband (Pico-I). 
This event was analysed by \citet{Ch}. The type of the source star was not determined, but they derived the LDC using a best fit parametric method to be $u_{LLD}=0.56\pm0.04$.  
%This was the only observation of this event that LDC was determined. 
Figure (\ref{fig:LC6}) shows the LD from the regularised FEM.
%Figure (\ref{fig:resultset-Data}) compares the LD profile from \citet{Ch} with our model-independent FEM. 

\textbf{MOA-2011-BLG-325/OGLE-2011-BLG-1101:}
This event was discovered and alerted by both MOA and OGLE surveys. The follow-up observations carried out by $\mu$FUN, PLANET, RoboNet and MiNDSTEp collaborations. 
% This event was analysed by \citet{Ch} for data sets from the other observatories.
We use the  Liverpool Telescope's I-passband observation (LT-I) has better data coverage during transit over the source star. Figure (\ref{fig:LC7}) shows the LD profile from the regularised FEM.

Finally, in order to check the consistency of our recovered profiles with standard LD profiles, we compare our recovered profiles (from regularized FEM) with the linear limb darkening (LLD) (see equation \ref{lld}) and the square root LD (see equation \ref{sqrt}) functions. Table (\ref{tab:LDCs}) represents the result of this comparison where we derive the parameters of these two LD profiles. We also compare our results with the LLD parameters that have been obtained from direct fitting to the light curves \citep{Ch,Ca06}. The result of our analysis from Table (\ref{tab:LDCs}) shows that events MOA-2010-BLG-436 and MOA-2011-BLG-093 are more consistent with the square-root profile than the linear limb darkening profile. The rest of events are consistent with the linear LD profile.

\section{Conclusion}
\label{conc}
In the finite-source effect during the microlensing events, the lens can transit over the source star for the small impact parameters. The intensity profile of the source star affect on the light curve around the peak on the microlensing events. Studying the light curve around the peak enables us to investigate the limb darkening (LD) profile of the source star. 

In this paper we used the classical Finite Element Method (FEM ) as an inversion tool to recover the 
LD equation from the observed microlensing light curve. We found out that this method has the two main problems 
of (i) instability with increasing the number of data points, (ii) a disperse recovered solution. In the second step we used the regularized FEM where we minimize an objective function that combines the residual of observed data with the reconstructed light curve as well as  minimizing dispersion of the intensity profile in the nodes, see equation (\ref{32}). This method could resolve the problems we faced in the classical FEM.

Finally we applied our method to single lens transit microlensing events and select data points from the light curve that has good coverage around the peak. We applied the regularized FEM to the following microlensing events of  OGLE-2004-BLG-254 (SAAO-I band), MOA-2007-BLG-233/OGLE-2007-BLG-302 (OGLE-I, MOA-R), MOA-2010-BLG-436 (MOA-R), MOA-2011-BLG-93 (Canopus) I-band, MOA-2011-BLG-300/OGLE-2011-BLG-0990 (Pico) I-band and MOA-2011-BLG-325/OGLE-2011-BLG-1101 (LT) I-Band and compare our model-independent results with that of other works where a simple model was assumed for LD of the star. We note that the advantage of this method would be the reconstruction of LD of source star without pre-assumption about the intensity profile of the source star. 

%Applying this method in the caustic crossing of binary events may reveal the profile of the source star with more accuracy.

 %features like spots and ellipticity of the source star. 

\section*{Acknowledgments}
We thank  C. Han and J. -Y. Choi for providing us the microlensing data.
We also thank M.A. Jalali and M. Dominik for useful discussion and sharing their insights with us. Also, we thank anonymous referee to his/her useful comments improving this work. 

%\nocite{*}

\label{lastpage}


\begin{thebibliography}{}

 \bibitem[\protect\citeauthoryear{Afonso et al.}{2003}]{Af2003}
 Afonso C. et al., 2003, A\&A, 400, 951 
  
%\bibitem[\protect\citeauthoryear{ Agol }{1996}]{Ag96}
 %Agol E., 1996, MNRAS, 279, 571
 
\bibitem[\protect\citeauthoryear{Albrow et al.}{1999}]{Al99} 
Albrow M. D. et al., 1999,  ApJ, 522, 1011. 

 \bibitem[\protect\citeauthoryear{Albrow et al.}{2001}]{Alb}
Albrow, M. D., et al. 2001, ApJ, 549, 759

  \bibitem[\protect\citeauthoryear{Alcock et al.}{1997}]{Al97}
 Alcock C. et al. 1997, ApJ, 491, 436

 \bibitem[\protect\citeauthoryear{Alcock et al.}{2000}]{Al2000}
 Alcock C. et al., 2000, ApJ, 542, 281

\bibitem[\protect\citeauthoryear{An et al.}{2002}]{An}
An J. H. et al., 2002, ApJ, 572, 521
 
 \bibitem[\protect\citeauthoryear{Aufdenberg et al.}{2006}]{Auf}
Aufdenberg J. P. et al., 2006, ApJ, 645, 664

\bibitem[\protect\citeauthoryear{Bogdanov \& Cherepashchuk }{1996}]{Bog}
 Bogdanov M. B., Cherepashchuk A. M., 1996, Astron. Rep., 40, 713

\bibitem[\protect\citeauthoryear{Burns et al.}{1997}]{Bu}   
 Burns D. et al., 1997, MNRAS, 290, L11

\bibitem[\protect\citeauthoryear{Cassan et al.}{2006}]{Ca06}
Cassan  A.  et al., 2006, A\& A, 460, 277

\bibitem[\protect\citeauthoryear{Cassan et al.}{2012}]{Ca12}
Cassan A. et al., 2012, Nature, 481, 167

\bibitem[\protect\citeauthoryear{Choi et al.}{2012}]{Ch}
Choi J. -Y. et al., 2012, ApJ, 751, 41

\bibitem[\protect\citeauthoryear{Chwolson}{1924}]{Ch24}
 Chwolson O., 1924, Astron. Nachrichten, 221, 329

\bibitem[\protect\citeauthoryear{Claret}{2000}]{Claret00}
Claret A., 2000, A\&A, 363, 1081
 
\bibitem[\protect\citeauthoryear{Craig \& Brown}{1986}]{CB86}    
Craig I. J. D., Brown J. C., 1986, Inverse Problems in Astronomy. Adam Hilger, Bristol

 \bibitem[\protect\citeauthoryear{Eddington}{1920}]{Ed20}
Eddington A. S., 1920, Space, Time and Gravitation. Cambridge Univ. Press, Cambridge

 \bibitem[\protect\citeauthoryear{Einstein}{1936}]{Ei36}
Einstein A., 1936, Science, 84, 506

 \bibitem[\protect\citeauthoryear{Fields et al.}{2003}]{Fie03}
Fields D. L. et al., 2003, ApJ, 596, 1305

 \bibitem[\protect\citeauthoryear{Gaudi}{2012}]{Ga12} 
Gaudi B. S., 2012, ARA\&A, 50, 411

 \bibitem[\protect\citeauthoryear{Gaudi \& Gould}{1999}]{Ga}
 Gaudi B. S., Gould A., 1999,  ApJ, 513, 619 
  
 \bibitem[\protect\citeauthoryear{Gaudi et al.}{2008}]{Ga08}  
Gaudi, B. S. et al., 2008, Science, 319, 927
    
 \bibitem[\protect\citeauthoryear{Golub \& Van Loan}{1996}]{GL}
Golub G. H., Van Loan C. E., 1996,  Matrix Computations.  Johns Hopkins Univ. Press, Baltimore

\bibitem[\protect\citeauthoryear{Gould}{1994}]{Go94}
Gould A., 1994, ApJ, 421, L71

\bibitem[\protect\citeauthoryear{Gray}{1992}]{Gr}  
Gray D. F., 1992, The Observation and Analysis of Stellar Photospheres. Cambridge Univ. Press, Cambridge

\bibitem[\protect\citeauthoryear{Gray}{2000}]{G00}
Gray N., 2000, astro-ph/0001359

\bibitem[\protect\citeauthoryear{Gray \& Coleman}{2000}]{GC01}
Gray N.,  Coleman I. J., 2001, in Menzies J. W., Sackett P. D., eds, ASP Conf. Ser. Vol. 239, Microlensing 2000: A New Era of Microlensing Astrophysics. Astron. Soc. Pac., San Francisco, p. 204
  
 \bibitem[\protect\citeauthoryear{Hendry, Bryce \& Valls-Gabaud}{2002}]{Hen02} 
  Hendry M. A., Bryce H. M., Valls-Gabaud D., 2002, MNRAS, 335, 539	%starspots

 \bibitem[\protect\citeauthoryear{Heyrovsk\'y}{2003}]{He}
  Heyrovsk\'y D., 2003, ApJ., 594, 464				%limb darkening
   
\bibitem[\protect\citeauthoryear{Heyrovsk\'y \& Sasselov}{2000}]{He2000}  
  Heyrovsk\'y D., Sasselov D., 2000, ApJ, 529, 69  %stellar spots
  
\bibitem[\protect\citeauthoryear{Jalali}{2010}]{Ja}  
Jalali M. A., 2010, MNRAS, 404, 1519		%FEM in astronomy

\bibitem[\protect\citeauthoryear{Jalali \& Tremaine}{2011}]{JT}
Jalali M. A., Tremaine S., 2011, MNRAS, 410, 2003

\bibitem[\protect\citeauthoryear{Mollerach \& Roulet}{2002}]{Mo}
 Mollerach S., Roulet E., 2002,  Gravitational Lensing and Microlensing. World Scientific Publishing Co. Pte. Ltd. Singapore 
 
 \bibitem[\protect\citeauthoryear{Moniez et al.}{2017}]{Mnz}  
Moniez M., Sajadian S., Karami M., Rahvar S., Ansari R., 2017, A\&A, 604, A124 %optical depth catalog analysis
  
\bibitem[\protect\citeauthoryear{Montargès et al.}{2014}]{Mnt}
Montargès M., Kervella P., Perrin G., Ohnaka K., Chiavassa A., Ridgway S. T., Lacour S., A\& A, 572, A17 % interferometry stellar photosphere 

\bibitem[\protect\citeauthoryear{Paczy\'nski}{1986}]{Paz1986} 
Paczy\'nski B., 1986, ApJ,  304, 1 							%microlensing by galactic halo
   
\bibitem[\protect\citeauthoryear{Paczy\'nski}{1996}]{Paz}
Paczy\'nski B., 1996, ARA\&A, 34, 419			%microlensing, local group

\bibitem[\protect\citeauthoryear{Perrin et al.}{2004}]{Per} 	%interferometry \alpha orionis
 Perrin G., Ridgway S. T., Coudé du Foresto V., Mennesson B., Traub W. A., Lacasse M. G., 2004, A\&A, 418, 675
    
\bibitem[\protect\citeauthoryear{Popper}{1984}]{Pop}   
 Popper D. M., 1984, AJ, 89, 1057					%eclipsing binaries
 
 \bibitem[\protect\citeauthoryear{Press et al.}{1992}]{Pr}
  Press W. H., Teukolsky S. A., Veltterling W. T., Flannery B. P., 1992,  Numerical Recipes in Fortran 77. Cambridge Univ. Press, Camberidge
  
  \bibitem[\protect\citeauthoryear{Rahvar}{2015}]{R15} 	%review
Rahvar S., 2015, IJMPD, 24, id.1530020
 
   \bibitem[\protect\citeauthoryear{Rahvar}{2016}]{Ra16} 		%SETI
 Rahvar S., 2016, ApJ, 828,19

  \bibitem[\protect\citeauthoryear{Rahvar \& Ghassemi }{2005}]{RG05} 	%microlensing astrometry
Rahvar S., Ghassemi S., 2005, A \& A, 438, L153.

\bibitem[\protect\citeauthoryear{Richichi \& Lisi}{1990}]{Ric}   		%lunar occultation, angular diameter of Antares
Richichi A., Lisi F., 1990, A\&A, 230, 355

\bibitem[\protect\citeauthoryear{Sajadian}{2015}]{S15}			%Polarimetry, stellar spots
Sajadian S., 2015, MNRAS, 452, 2587

\bibitem[\protect\citeauthoryear{Sajadian \& Rahvar}{2015}]{SR15}	%Photometric, astrometric and polarimetric
Sajadian S., Rahvar S., 2015, MNRAS, 452, 2579

\bibitem[\protect\citeauthoryear{Schneider \& Weiss}{1986}]{sch86} 	%binary lens
Schneider P., Weiss A., 1986, A\&A, 164, 237

\bibitem[\protect\citeauthoryear{Schneider \& Wagoner }{1987}]{sch87} 	%Amplification and Polarization of Supernovae by Gravitational Lensing
Schneider P., Wagoner R. V., 1987, ApJ, 314, 154

\bibitem[\protect\citeauthoryear{Simmons, Willis \& Newsam}{1995}]{SWN95}	%first paper on polarimetry in microlensing
 Simmons J. F. L.,  Willis J. P., Newsam A. M., 1995, A\&A, 293, L46 
 
\bibitem[\protect\citeauthoryear{Southworth et al.}{2005}]{Sou}   		% Eclipsing binaries, radii ...
Southworth J., Smalley B., Maxted P. F. L., Claret  A., Etzel P. B., 2005, MNRAS, 363, 529

 \bibitem[\protect\citeauthoryear{Southworth et al.}{2015}]{Sou2}  %Transits in the WASP-57 planetary system		
Southworth, J. et al., 2015, MNRAS, 447, 711

\bibitem[\protect\citeauthoryear{Tsapras}{2018}]{Ts18}			%review on extrasolar planets
Tsapras Y., 2018, Geosciences, 8, 365

 \bibitem[\protect\citeauthoryear{Valls-Gabaud}{1995}]{DV95} 
Valls-Gabaud D., 1995, in Muecket J. P., Gottloeber S., Mueller V., eds, Large scale structure in the universe. World Scientific Co., Potsdam, Germany, p.326

 \bibitem[\protect\citeauthoryear{Valls-Gabaud}{1998}]{DV98}
Valls-Gabaud D., 1998, MNRAS, 294, 747
  
 \bibitem[\protect\citeauthoryear{Valls‐Gabaud}{2006}]{DV} 
Valls-Gabaud D., 2006, in Alimi J. M., Füzfa A., eds, AIP Conf. Vol. 861, Albert Einstein Century International Conference, LUTH, Paris, p. 1163
 
\bibitem[\protect\citeauthoryear{Wazwaz}{2011}]{Waz11}
Wazwaz A.M., 2011, Linear and Nonlinear Integral Equations: Methods and Applications. Higher education press, Beijing and Springer-Verlag Berlin Heidelberg

\bibitem[\protect\citeauthoryear{Walker}{1995}]{Wa95}
Walker M.A., 1995,  ApJ, 453, 37

\bibitem[\protect\citeauthoryear{Witt}{1995}]{Wi95}
Witt H. J., 1995, ApJ., 449, 42.

\bibitem[\protect\citeauthoryear{Witt \& Mao}{1994}]{Wi94}
Witt H. J., Mao S., 1994,  ApJ., 430, 505

\bibitem[\protect\citeauthoryear{Wittkowski et al.}{2006}]{Wit}		%Tests of stellar model atmospheres by optical interferometry
Wittkowski M., Aufdenberg J. P., Driebe T., Roccatagliata V., Szeifert T., Wolff B., 2006, A\&A, 460, 855

\bibitem[\protect\citeauthoryear{Yoo et al.}{2004}]{Yoo}		%Finite-Source Effects from a Point-Mass Lens
Yoo J. et al. 2004, ApJ, 603, 139

\bibitem[\protect\citeauthoryear{Zienkiewicz, Taylor \& Zhu}{2005}]{Zie}
Zienkiewicz O. C., Taylor R. L., Zhu J. Z., 2005, The Finite Element Method:its Basis and Fundamentals. Elsevier, Butterworth-Heinemann, Oxford

\bibitem[\protect\citeauthoryear{Zub et al.}{2011}]{Zub} 		%Limb-darkening measurements for a cool red giant
Zub M. et al., 2011, A\&A, 525, A15

\end{thebibliography}
\end{document}